\documentclass[times,preprint,review]{elsarticle}

\usepackage{graphicx}
\graphicspath{ {./images/} }
\usepackage{subcaption}
\usepackage{amssymb}
\usepackage{latexsym}

\usepackage{lineno}

\newcommand\wf{0.95}

\begin{document}


\begin{frontmatter}


\title{Image Processing and Quality Control for Abdominal Magnetic Resonance Imaging in the UK Biobank}



\author[1]{Nicolas Basty}
\ead{n.basty@westminster.ac.uk}
\author[2]{Yi Liu}
\author[2]{Madeleine Cule}
\author[1]{E. Louise Thomas}
\author[1]{Jimmy D. Bell}
\author[1]{Brandon Whitcher\corref{cor1}}
\ead{b.whitcher@westminster.ac.uk}
\cortext[cor1]{Corresponding author.}

\address[1]{Research Centre for Optimal Health, School of Life Sciences, University of Westminster, London, UK}
\address[2]{Calico Life Sciences LLC, South San Francisco, California, USA}

\begin{abstract}

An end-to-end image analysis pipeline is presented for the abdominal MRI protocol used in the UK Biobank on the first 38,971 participants.  Emphasis is on the processing steps necessary to ensure a high-level of data quality and consistency is produced in order to prepare the datasets for downstream quantitative analysis, such as segmentation and parameter estimation.  Quality control procedures have been incorporated to detect and, where possible, correct issues in the raw data.  Detection of fat-water swaps in the Dixon series is performed by a deep learning model and corrected automatically.  Bone joints are predicted using a hybrid atlas-based registration and deep learning model for the shoulders, hips and knees.  Simultaneous estimation of proton density fat fraction and transverse relaxivity (R2*) is performed using both the magnitude and phase information for the single-slice multiecho series.  Approximately 98.1\% of the two-point Dixon acquisitions were successfully processed and passed quality control, with 99.98\% of the high-resolution T1-weighted 3D~volumes succeeding.  Approximately 99.98\% of the single-slice multiecho acquisitions covering the liver were successfully processed and passed quality control, with 97.6\% of the single-slice multiecho acquisitions covering the pancreas succeeding.  At least one fat-water swap was detected in 1.8\% of participants.  With respect to the bone joints, approximately 3.3\% of participants were missing at least one knee joint and 0.8\% were missing at least one shoulder joint.  For the participants who received both single-slice multiecho acquisition protocols for the liver a systematic difference between the two protocols was identified and modeled using multiple linear regression.  The findings presented here will be invaluable for scientists who seek to use image-derived phenotypes from the abdominal MRI protocol.

\end{abstract}


\end{frontmatter}


\section{Introduction}

The aim of the UK Biobank (UKBB) project, started in 2006, has been to gather lifestyle information, biometric, and genetic data for 500,000 individuals in the UK population for research purposes \citep{sudlow2015uk}.  In 2014, the project was expanded to include imaging for 100,000 of the participants, which will create the largest and most extensive collection of medical imaging data in the world, once completed \citep{littlejohns2020biobank}.  Rigorous and standardized multimodal scanning protocols, including a variety of acquisition protocols, are being performed seven days a week in four sites across the UK (Stockport, Newcastle, Reading, and Bristol).  Volunteers from the initial UKBB cohort aged 40 to 69 years have been included in the study.  The in-depth phenotyping protocol includes several imaging modalities, providing a comprehensive set of structural and functional data for every participant that includes: dual-energy X-ray absorptiometry, neck and carotid artery ultrasound, as well as brain, cardiac, and abdominal magnetic resonance imaging (MRI).  In terms of MRI scanning, each region of the body has a specific time allocation: brain 30 minutes, heart 20 minutes and abdomen 10 minutes, with three subjects rotating between measurements at the UKBB scanning site at the same time to achieve the daily target of 17 subjects.  If the entire set of scans for a particular protocol is not acquired within the allocated time, the radiographers have been instructed to move on to the next stage of the protocol, and consequently some subjects may have incomplete datasets. 

The quality and reproducibility of any MRI scanning protocol is dependent on a number of factors including the skills of the radiographer (e.g., subject placement, subject alignment, slice position), on the presence of imaging artifacts (e.g., caused by subject motion, motion due to breathing or magnetic-field inhomogeneities) and the time limits imposed for the actual acquisition.  In the case of the Abdominal Protocol, limited to 10 minutes, it meant that the relative quality of the images (e.g., signal-to-noise) was partly compromised in order to maximize the overall coverage and variety of acquisitions. This protocol includes a two-point 3D Dixon sequence covering the body between the neck and the knees, a high-resolution 3D T1-weighted (T1w) pancreas acquisition, as well as two quantitative single-slice acquisitions that focus on the liver and the pancreas, respectively.  The Dixon protocol involves six separate acquisitions and requires extensive preprocessing before subsequent analyses can be performed.  This type of acquisition is prone to swaps when separating the fat and water channels during reconstruction by the scanner software, especially in structures that are physically separated (e.g., arms, legs) in the different series \citep{ma2008dixon}.  With the high throughput of subjects performed under a tight time schedule, mishaps will inevitably occur at certain points in the protocol.  Given these potential issues, and being an extremely large database covering a heterogeneous set of organs, structures and tissues, the Abdominal Protocol of the UK Biobank demands a very robust and thorough image analysis pipeline with several layers of quality control.  Such a pipeline ensures that consistent downstream analysis of the acquired data can be achieved with minimal manual intervention.  With 100,000 datasets to be produced, relying solely on visual inspection for quality control is impractical.  

Over the last decade, MRI has become the gold standard for body composition, particularly when measuring adipose tissue, liver and pancreatic fat content. Some of these measurements have had an enormous impact on our understanding of metabolic conditions such as type-2 diabetes and nonalcoholic fatty liver disease \citep{thomas2013whole}. In addition to these measurements, the Abdominal Protocol includes multiple tissues and organs such as muscles, kidneys, spleen, vertebra, lungs, pelvis, femur, etc, with the potential for a myriad of clinically relevant variables. It is therefore somewhat surprising that to date there have been relatively few publications arising from the Abdominal Protocol (13), compared to those arising from the brain (81) and cardiac (52) protocols\footnote{The results come from individual Pubmed searches using the keywords: ``UK Biobank'', ``MRI'' and either ``abdominal'', ``brain'' or ``cardiac''.}. This reflects the paucity of abdominal  image-derived phenotypes that have been returned to the UK Biobank, which in turn reflects the lack of robust automated methods for a comprehensive analysis of the abdominal MR images.

The studies published so far arising from the Abdominal Protocol, including those utilising the single-slice multiecho liver data, have performed quality control by visual inspection with manual annotations necessary for quantification of parameters \citep{wilman2017characterisation, mckay2018measurement, mojtahed2019reference, wojciechowska2018automated, triay2019magnitude}. Similarly, the neck-to-knee Dixon data have been used in several studies investigating body composition \citep{west2016feasibility, linge2019sub, linge2018body, borga2018advanced} as well as predictive models for biometry and markers of age \citep{langner2019identifying, langner2020large}.  These studies do not appear to include automated quality control procedures, with some requiring manual editing of erroneous data and all of them referring to visual inspection for quality control of the results.  With the UK Biobank expecting to complete scanning the cohort of 100,000 subjects by 2023, and a significant number of participants due to undergo an additional follow-up scan, relying on manual expert work or visual inspection for a collection of data at this scale is not feasible, particularly for the three-dimensional data, which contain several orders of magnitude more information compared with single-slice acquisitions.  

Once published, the UK Biobank expects researchers to submit their quantitative results for any derived biomarkers in order to further enrich the database and accelerate research.  With momentum building on previous studies and their results \citep{ji2019genome, van2020impact, wilman2019genetic}, the community clearly will benefit from extensive quality control.  Preprocessing and quality control using automation where possible has been performed using the first 10,000 subjects in brain imaging \citep{alfaro2018image} and the first 19,265 subjects for cardiac MRI \citep{tarroni2020large}, reporting on the high quality of the data using machine learning to quantify motion-induced misalignment, quality of the contrast, and short-axis stack coverage of the heart.  These quality control evaluations report an overall high quality of data, with 81.06\% of the brain data having all of the acquisitions and estimated to be used for analysis.  For the cardiac MRI, over 99.9\% of the data are complete, with 85.8\% having optimal coverage, 84.0\% not suffering from major misalignment and 97.9\% with good image contrast.  For the Abdominal Protocol, \citet{west2016feasibility} presented detailed quality control results for the Dixon acquisition in the first 3,000 subjects, but required visual inspection of the images.  

In this paper we present a comprehensive pipeline for the automatic preprocessing and quality control of the UKBB Abdominal Protocol, as well as a summary of the data for the first 38,971 subjects.

\section{Materials and Methods}

\subsection{Data Acquisition}
\label{sec:data_acquisition}

Analysis was performed on all available datasets as of December 2019, with 38,971 abdominal MRI datasets released by the UK Biobank, where a total of 100,000 datasets are the ultimate goal for the imaging sub-study.  All of the abdominal scans were performed using a Siemens Aera 1.5~T scanner (Syngo MR D13) (Siemens, Erlangen, Germany).  Acquisition parameters for the Abdominal Protocol have been provided in \citet{littlejohns2020biobank}.  We focus here on four separate acquisitions, with one sequence being applied twice (once for the liver and once for the pancreas):
\begin{enumerate}
    \item The Dixon sequence involved six overlapping series that were acquired using a common set of parameters: TR = 6.67~ms, TE = 2.39/4.77~ms, FA = $10^\circ$ and bandwidth = 440~Hz.  The first series, over the neck, consisted of 64~slices, voxel size $2.232 \times 2.232 \times 3.0$~mm and $224 \times 168$ matrix; series two to four (covering the chest, abdomen and pelvis) were acquired during 17 sec expiration breath holds with 44~slices, voxel size $2.232 \times 2.232 \times 4.5$~mm and $224 \times 174$ matrix; series five, covering the upper thighs, consisted of 72 slices, voxel size $2.232 \times 2.232 \times 3.5$~mm and $224 \times 162$ matrix; series six, covering the lower thighs and knees, consisted of 64~slices, voxel size $2.232 \times 2.232 \times 4$~mm and $224 \times 156$ matrix.
    \item The high-resolution T1w acquisition sequence for the pancreas was acquired under a single expiration breath hold with TR = 3.11~ms, TE = 1.15~ms, FA = $10^\circ$, bandwidth = 650~Hz, voxel size $1.1875 \times 1.1875 \times 1.6$~mm and $320 \times 260$ matrix.  
    \item The single-slice multiecho gradient echo acquisition sequence, for the liver and pancreas, was acquired using the common set of parameters: TR = 27~ms, TE = 2.38/\-4.76/\-7.15/\-9.53/\-11.91/\-14.29/\-16.67/\-19.06/\-21.44/\-23.82~ms, FA = $20^\circ$, bandwidth = 710~Hz, voxel size $2.5 \times 2.5 \times 6.0$~mm and $160 \times 160$ matrix.  This was implemented for the pancreas only after the first approximately 10,000 subjects.
    \item The single-slice IDEAL sequence \citep{reeder2005iterative} for the liver used the following parameters: TR = 14~ms, TE = 1.2/\-3.2/\-5.2/\-7.2/\-9.2/\-11.2~ms, FA = $5^\circ$, bandwidth = 1565~Hz, voxel size $1.719 \times 1.719 \times 10.0$~mm and $256 \times 232$ matrix.  This replaced the single-slice multiecho sequence for the liver after the first approximately 10,000 subjects.
\end{enumerate}

\noindent The North West Multicentre Research Ethics Committee (MREC), UK, approved the study and written informed consent was obtained from all subjects prior to study entry.

\subsection{Image Analysis Pipelines}

All image processing and quantitative analysis has been performed in Python~3 \citep{python2009} using both CPU and GPU computing resources.  

\subsubsection{Data conversion} 

\begin{figure}[t]
  \centering
  \includegraphics[width=\wf\textwidth]{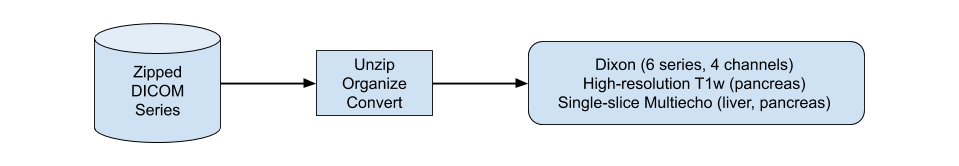}
  \caption{Workflow for data conversion and data management.  Data from the Abdominal Protocol are downloaded as individual zip files for each series.  These files are uncompressed, organized and converted into multidimensional (3D or 4D) arrays in the NIfTI-1 data format.}
  \label{fig:dicom_to_nifti}
\end{figure}

All DICOM files are converted to the NIfTI-1 format using the \texttt{dcmstack}\footnote{https://github.com/moloney/dcmstack} python package (Figure~\ref{fig:dicom_to_nifti}), stacking multiple DICOM files into a single multidimensional NIfTI-1 object, containing the DICOM metadata.  Preserving the DICOM metadata is important for downstream processing steps; e.g., assembling multiple series into a single neck-to-knee volume for the Dixon acquisition or parameter estimation in the multiecho acquisitions.

\subsubsection{Data Management}

\begin{figure}[tbp]
  \centering
  \includegraphics[width=\wf\textwidth]{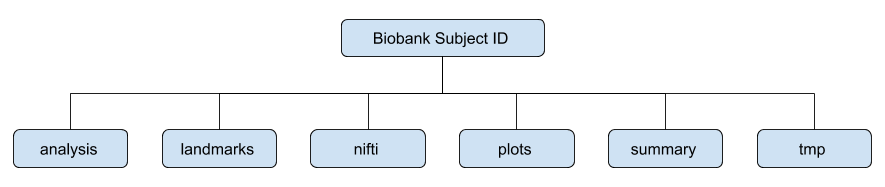}
  \caption{Directory tree structure for a single abdominal MR scanning session.}
  \label{fig:directory_tree}
\end{figure}

The following fixed directory tree structure were used in order to streamline image analysis at scale (Figure~\ref{fig:directory_tree}).
\begin{itemize}
    \item \texttt{analysis}: NIfTI files derived from the original datasets; e.g., binary masks and parameter estimates.
    \item \texttt{landmarks}: the predicted landmarks from the bone joints (cf. Section~\ref{sec:bone_joints}).  
    \item \texttt{nifti}: NIfTI files associated with the four acquisition protocols, after being processed by their respective image analysis pipelines.  The number of files depended on the number of MR sequences successfully acquired in the scanning session.
    \item \texttt{plots}: PNG files that captured results from intermediate steps in the pipeline.
    \item \texttt{summary}: summaries of the image data, processing steps and parameter estimates.
    \item \texttt{tmp}: files useful for debugging and troubleshooting in development that were deleted prior to archival, including the original DICOM files.  
\end{itemize}
Fixed file names were also defined to simplify automation.  

\begin{figure}[tbp]
  \centering
  \includegraphics[width=\wf\textwidth]{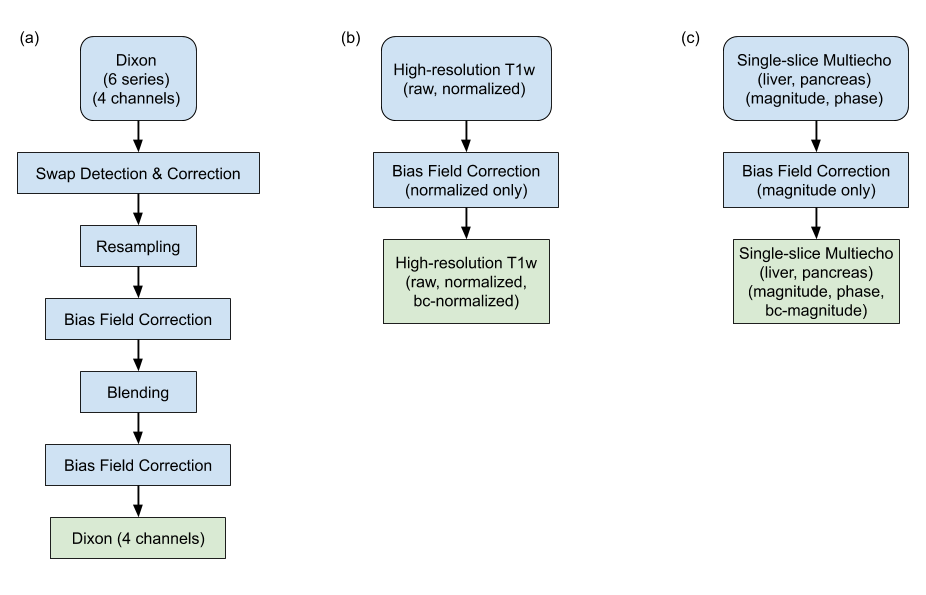}
  \caption{Workflows for the distinct acquisition protocols: (a) Dixon data involve a total of 24 files (four channels and six series) that go through several steps to produce high-quality neck-to-knee image volumes, (b) Bias field correction is applied to the single high-resolution T1w series dedicated to the pancreas and (c) Bias field correction is applied to the single-slice multiecho acquisitions for both the pancreas and liver.}
  \label{fig:series_pipelines}
\end{figure}

\subsubsection{Dixon Pipeline}
\label{sec:dixon_pipeline}

The six separate series associated with the two-point Dixon acquisition were positioned automatically after the initial location was selected by the radiographer \citep{west2016feasibility,littlejohns2020biobank}.  Reconstruction of the fat and water channels from the two-point Dixon acquisition was performed on the scanner console.  Four sets of DICOM files were generated for each of the six series in the neck-to-knee Dixon protocol: in-phase, opposed-phase, fat and water.  The preprocessing steps are summarized in Figure~\ref{fig:series_pipelines}(a). 

\begin{figure}[tbp]
  \centering
  \includegraphics[width=\wf\textwidth]{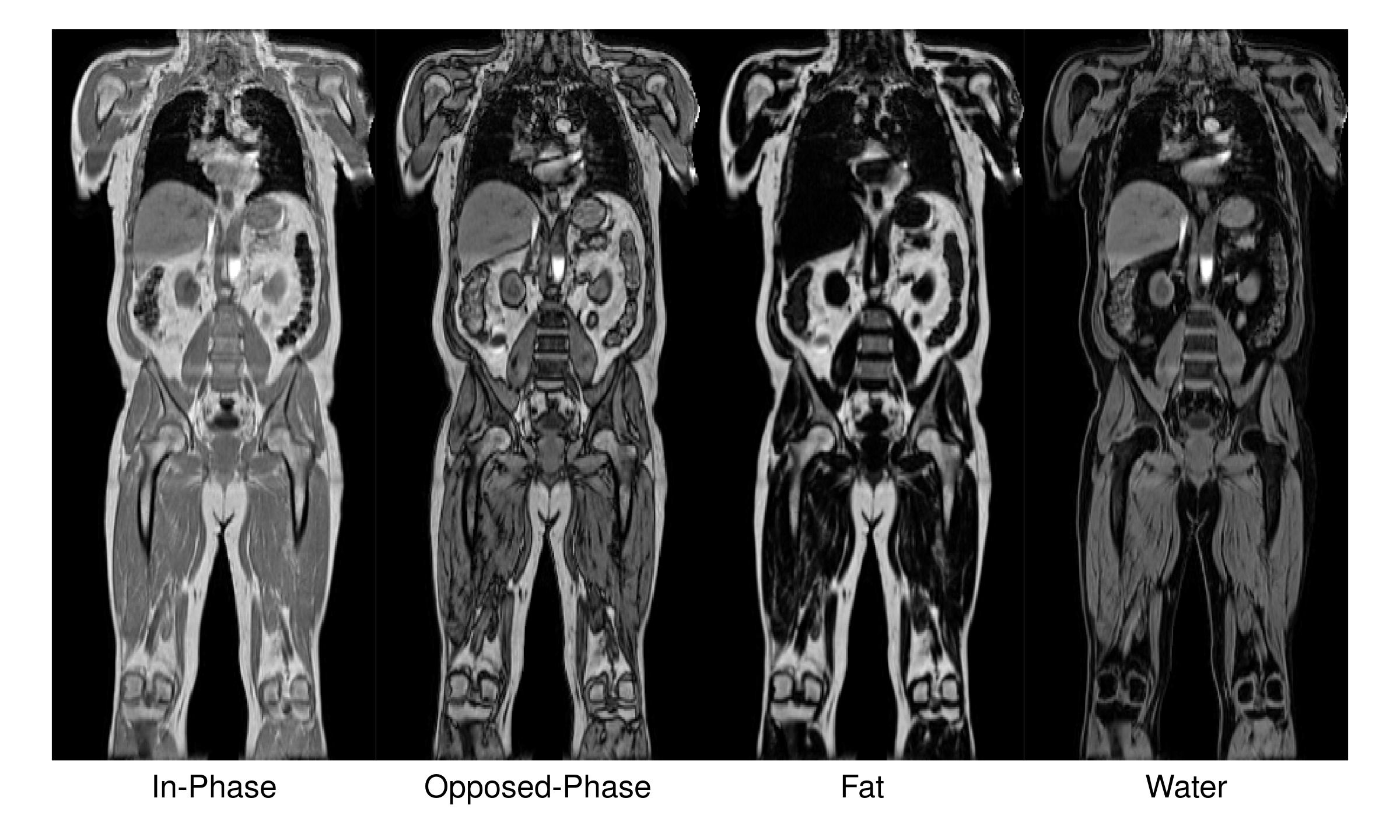}
  \caption{Complete neck-to-knee Dixon data after successful preprocessing via the image analysis pipeline for a single subject.  The four channels are (from left-to-right):  in-phase, opposed-phase, fat and water.}
  \label{fig:four_channels}
\end{figure}

Bias field correction \citep{tustison2010n4itk} was performed on the in-phase volume and the resulting bias field applied to the other channels (opposed-phase, fat, water) for each series.  The series were resampled to a single dimension and resolution to facilitate merging of all six series into a single three-dimensional volume (size = [224, 174, 370], voxel = $2.232 \times 2.232 \times 3.0$~mm).  To reduce the effect of signal loss when blending the series, we identified the fixed set of slices that form an overlap (inferior-superior direction) between adjacent series.  We applied a nonlinear function to blend the signal intensities on these regions of overlap, where slices in the interior of the volume were heavily weighted and slices near the boundary were suppressed.  We repeated the bias field correction on the blended in-phase volume and applied the estimated bias field to the other channels.  The final output from a representative Dixon acquisition is displayed in Figure~\ref{fig:four_channels} for a single subject.

\begin{figure}[tbp]
  \centering
  \includegraphics[width=\wf\textwidth]{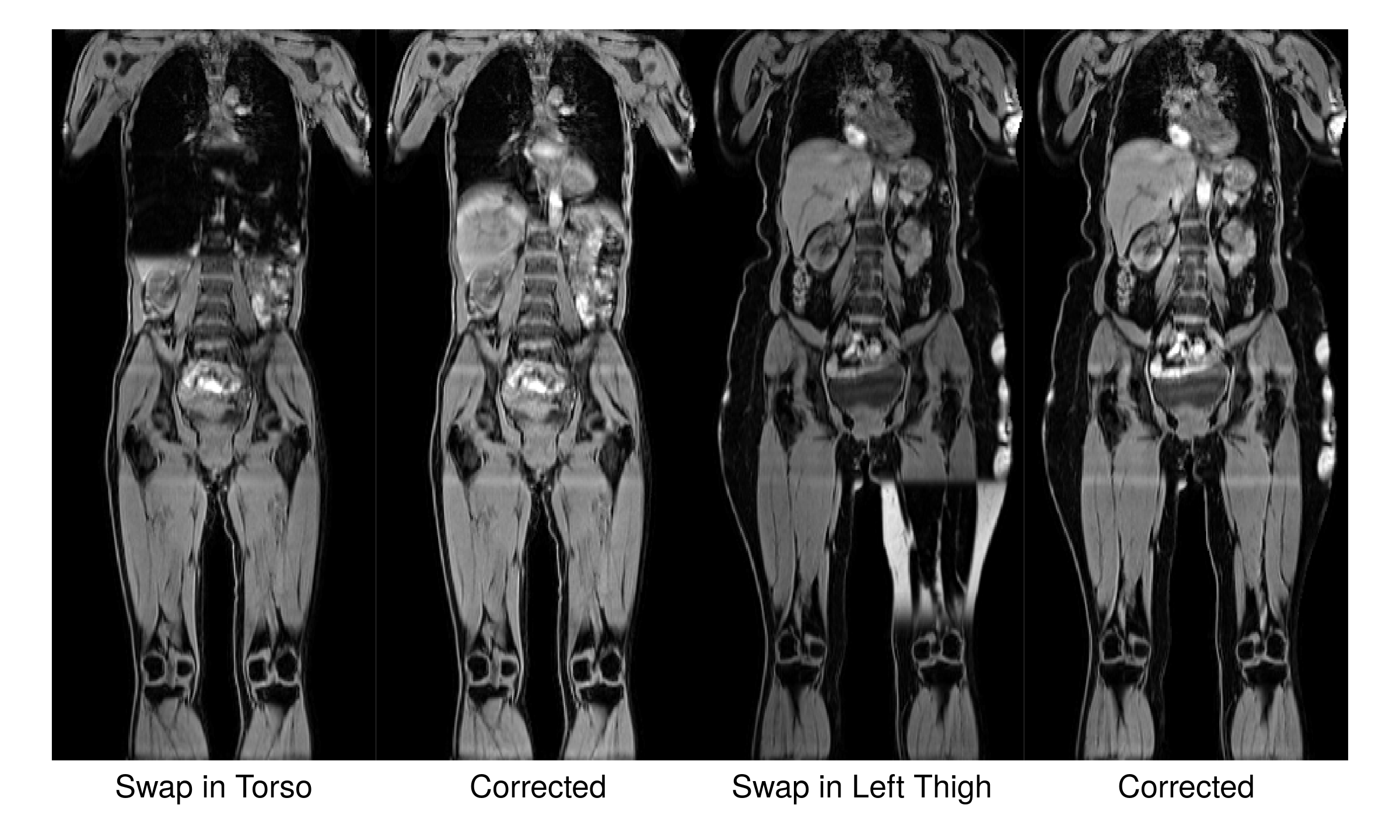}
  \caption{Examples of fat-water swaps in the UK Biobank.  The water channel is displayed in all images.  There is a swap in the second series (chest) of the subject on the left, and there is a swap in one of the legs in the fifth series (right thigh) of the subject on the right.}
  \label{fig:swap_examples}
\end{figure}

Fat-water swaps are a common issue in the reconstruction of Dixon acquisitions, where the fat and water labels attributed to the reconstructed images are reversed for all voxels in the acquired data series or cluster of voxels associated with separate anatomical structures (e.g., legs or arms).  Examples are provided in Figure~\ref{fig:swap_examples}, where the water channel for the subject on the left suffers from a swap in the second series (in the torso) and the fat channel is labeled as water (not shown).  The water channel for the subject on the right suffers from a swap in the fifth series (in the left thigh) and again the fat channel is incorrectly labeled as water (not shown).  Once corrected, the water channel for both subjects is consistent.  We used a convolutional neural network (CNN) model to detect swaps, with six individual models trained for each of the six acquired series.  Only fat-water swaps that involved the entire series or the left-right halves in the final two series were considered.  Partial fat-water swaps (e.g., the top of the liver) are an area of future work. 

Each model used a sequential architecture with six layers that assigned a label, zero for water and one for fat, to each of the series when given a central 2D slice from the series. Each convolution block ($C_n$) was made up of $n$ convolutions that were $3 \times 3$ spatial filters applied with stride of length two, followed by a leaky rectified linear unit (ReLU) activation with slope 0.2 and batch normalization. The final layer had stride of length three and a sigmoid activation for binary classification of the input as either water or fat. The number of convolution filters was doubled in each layer down the network as follows: $C_{64} \rightarrow C_{128} \rightarrow C_{256} \rightarrow C_{512} \rightarrow C_{1024} \rightarrow C_{1}$. One model using the same architecture was trained for each series. The two models covering the bottom two series that include the legs checked the right and left half of the input image separately to accommodate for the legs being separate structures with increased likelihood of swaps. Each of the six series for 462 subjects were individually inspected to ensure no swaps occurred and used to train the models. The ten central coronal slices of each subject were selected by checking the image profile of the slice in each slab, where the largest profile was assumed to be the centre of the body. Thus, a total of 4620 images were available for training each of the networks. No additional data augmentation was performed. Each 2D slice was normalized and cropped by 32 voxels at both (left, right) ends to exclude the boundaries of the field of view and minimize the influence of voxels outside the body and in the arms. The model was trained with a binary cross entropy loss function using the Adam optimizer and a batch size of 100 until convergence, which was between 150 and 200 epochs depending on the series. The models were validated on a separate set of 615 subjects, resulting in 4,920 individual swap detection operations performed as every set of Dixon data is subject to eight classification tests. The validation, via visual inspection of all the series and the swap detection results, revealed only two instances of the second series (the chest) and one instance of the fifth series (one of the two legs) were mislabeled out of the total 4,920 checks performed. Only one false positive, in the second series, was observed.

\subsubsection{T1-weighted Pipeline}

\begin{figure}[tbp]
  \centering
  \includegraphics[width=\wf\textwidth]{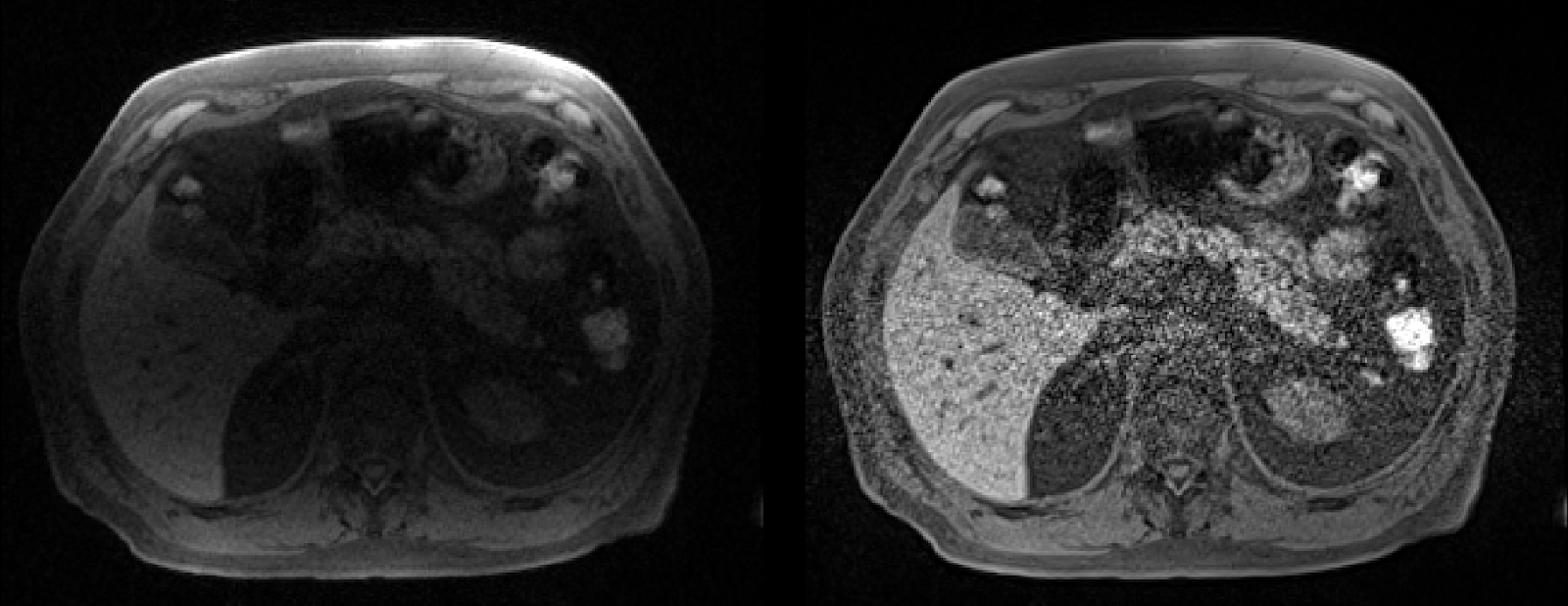}
  \caption{Mid-axial slice from the high-resolution T1-weighted 3D volume centered on the pancreas.  Raw (left) and normalized (right) data are shown.}
  \label{fig:t1w_pancreas}
\end{figure}

A single 3D T1-weighted (T1w) volume of higher-resolution data was acquired, centered on the pancreas.  Two versions were provided, with and without normalization, from the acquisition (Figure \ref{fig:t1w_pancreas}).  Bias field correction was performed to reduce signal inhomogeneities in the normalized volume (cf. Figure~\ref{fig:series_pipelines}(b)).  No additional preprocessing was applied to the high-resolution 3D T1w pancreas volumes.

\subsubsection{Single-slice Multiecho Pipeline}
\label{sec:single_slice_multiecho}

Two single-slice multiecho acquisitions were performed, one positioned through the liver (orthogonal orientation) and another one positioned through the pancreas (oblique orientation).  The liver acquisition parameters were changed after the pilot scanning period and replaced with the IDEAL acquisition \citep{reeder2005iterative}.  We performed bias field correction on each echo time of the magnitude images to overcome signal inhomogeneities.  For organ/tissue segmentation, we used only the bias-field corrected magnitude images.  We used the unprocessed data, both magnitude and phase, to estimate fat fraction and transverse relaxivity.


The limitations of two-point Dixon acquisitions for fat quantification have been well documented in the literature \citep{bydder2008relaxation, yu2008multiecho}.  Software\footnote{https://github.com/marcsous/pdff} available from Dr~Mark Bydder, specifically the PRESCO (Phase Regularized Estimation using Smoothing and Constrained Optimization) algorithm \citep{bydder2020constraints}, was used to simultaneously estimate the proton density fat fraction (PDFF) and transverse relaxivity (R2*) values voxelwise from the single-slice multiecho gradient echo (GRE) and IDEAL acquisitions.  Essentially, a multi-peak spectrum was constructed from the echo times in the acquisition protocol and used to perform nonlinear least squares under multiple regularization constraints that extends the IDEAL (Iterative Decomposition of Water and Fat with Echo Asymmetry and Least-Squares Estimation) algorithm \citep{reeder2005iterative, yu2008multiecho}.


It is also possible to convert transverse relaxivity (R2*) into iron concentration (in mg/g) using a linear transformation.  Within the literature a number of different formulas have been proposed, for consistency we adopted the same approach that has previously been applied to data from the UK Biobank in \citet{mckay2018measurement}, where 
\begin{equation}
\label{eqn:iron_concentration}
    \mathrm{iron~concentration} = 0.202 + 0.0254 \cdot \mathrm{R2^*}.
\end{equation}
Equation~\ref{eqn:iron_concentration} was originally reported in \citet{wood2005mri}.

\subsection{Quality Control}



\subsubsection{Anomaly Detection}

Anomaly detection of the final reconstructed volumes was performed to identify potential data issues such as image artifacts, positioning errors or missing series.  This was achieved via measurement of the dimensions from the final reconstructed volume and edge detection performed on the binary body mask.  To generate the body mask, we applied multi-scale adaptive thresholding to the flattened in-phase signal intensities, keeping only the largest connected component, then performed a binary closing operation.  The presence of sharp edges in the body mask highlighted discontinuities in the data and was used as an indicator of data inconsistencies.  We used Canny edge detection on a central coronal slice and a sagittal slice of the body mask containing both background and subject labels. In a normal subject, edge detection should not highlight anything other than the vertical contour of the body from neck to knee.  Presence of discontinuities or horizontal features in the body mask were indicators of anomalies. Clusters of voxels in the edge image corresponding to horizontal edges exceeding a threshold 10 voxels in the sagittal and coronal slice, or 25 in either slice, triggered the anomaly detection. Those values were selected based on results of 1,000 subjects. Field of view errors in positioning the subject were identified if the head or chin were partly or fully visible, or if the total volume did not match the standard $224 \times 174 \times 370$ dimension of the correctly assembled Dixon acquisition.  Signal dropout artifacts were caused by metal objects such as knee or tooth implants and identified when discontinuities appeared inside the body mask. 

\subsubsection{Bone Joint Detection}
\label{sec:bone_joints}

Shoulder, hip, and knee landmarks are useful for quality control as well as downstream analysis.  For example, the absence of anatomical landmarks such as the shoulders or the knees would highlight an issue with the data acquisition, such as sub-optimal placement of the subject in the field of view, or very tall subjects where neck-to-knee coverage is not possible due to the fixed number of series acquired.  The bone joint landmarks were defined to be the centroid of the head of the humerus for the shoulders, the centroid of the head of the femur for the hips, and the center of the femur just above where the medial and lateral condyles form for the knees.  Six bone joints ($\mathrm{shoulder}~\times~2$, $\mathrm{hip}~\times~2$, $\mathrm{knee}~\times~2$) were manually placed for 197 subjects and used to train the models.  When predicting the bone joints for a new subject an initial guess of the bone joint coordinates was made using atlas-based registration with a subset of 35 subjects chosen to match gender and ranked by height \citep{aljabar2009multi}, taking the centroid from the cluster of co-registered points at each anatomical location.

\begin{figure}[tbp]
  \centering
  \includegraphics[trim={1cm 2.5cm 1cm 2.5cm},clip,width=\wf\textwidth]{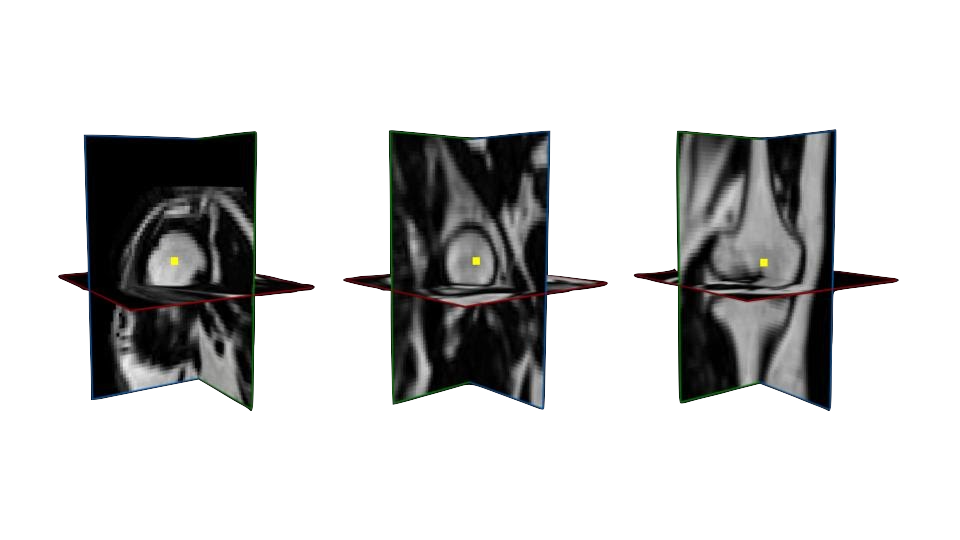}
  \caption{Example training data for the automated detection of bone joints.  Single voxels were placed in the centroid of the left/right humerus head (shoulders), left/right femur head (hips) and area just above where the medial and lateral condyles form in the left/right femurs (knees).}
  \label{fig:bone_joints_input}
\end{figure}

The centroid of the coordinates from the atlas-based registrations was used to define a $64 \times 64 \times 64$ cube around the ground-truth bone joint coordinates and used as input to the model (Figure~\ref{fig:bone_joints_input}).  Data from the left and right structures were combined so that a single model was trained for each bone joint, exploiting the symmetry of those structures in the human body.  The model architecture may be described as a $2 \times 2 \times 2$ average pooling followed by three pairs of convolutional layers with filter sizes $3\times 3 \times 3$, followed by a $2 \times 2 \times 2$ maximum pooling operation and increasing number of filters pairwise from 32 to 128.  Batch normalization and ReLU activations were used throughout and a final fully-connected layer with a leaky ReLU activation producing the vector of predicted three-dimensional coordinates. Each model was trained over 500 epochs using the Adam optimizer minimizing a mean squared error loss function.  For each individual bone joint if less than 50\% of the co-registered coordinates were in the field of view, the procedure did not apply the predictive model and recorded that bone joint as missing.  To validate the predictive model a modest set of test data (48 volumes, 20\% of the total number) was held back from the training step.  

\subsubsection{Anatomical Coverage of the Single-Slice Acquisitions}
\label{sec:slice_location}

As discussed in Section~\ref{sec:single_slice_multiecho}, additional single-slice multiecho acquisitions were performed in the liver and pancreas.  The multiecho slices were placed manually by the radiographers using information from the Dixon acquisition to facilitate correct anatomical coverage.  The liver slice was orthogonal and equivalent to an axial slice through the body at the height of the liver.  To identify sequences with poor slice placement, we resampled the single-slice data into the Dixon acquisition in order to obtain its equivalent slice location in that image space.  The anatomical coverage is only approximate since the two acquisitions (single-slice multiecho and Dixon) were performed on separate breath holds.

The pancreas has a much more complicated shape and position than the liver, and an oblique axial slice was placed by the radiographer in order to maximize anatomical coverage for subsequent analysis.  Given the oblique placement a straightforward assessment using slice location was not possible (as compared to the liver).  Hence, after resampling the single-slice data into the high-resolution T1w acquisition a census on the number of voxels that comprise the resulting 2D~mask was calculated to provide an initial assessment of the anatomical coverage of the slice placement on the pancreas.  Two-dimensional masks of the pancreas were available to evaluate slice placement using the method described in \citet{basty2020automated}. 

\subsection{Statistical Analysis}

All regression models were fit using the \textsf{R} software environment for statistical computing and graphics \citep{r2020}.

\section{Results}

\subsection{Data Conversion}


As specified in Section~\ref{sec:data_acquisition} there were up to five distinct acquisitions that were accommodated in the current image analysis pipeline.  A total of 38,935 individuals had at least one successful acquisition; the breakdown for each data type, and the number of successful analysis ready datasets, is given in Table~\ref{tab:numbers}.  The Dixon acquisition had the highest number of available datasets, and the Dixon pipeline involved multiple steps that go beyond simply converting DICOM files to the NIfTI-1 format (cf. Section~\ref{sec:dixon_pipeline}).  Of the 234 scans, representing 0.6\% of the total number of datasets available, that did not complete the pipeline:

\begin{itemize}
    \item There were 74 non-standard acquisitions with additional series being acquired beyond the expected six series.  These datasets did contain the six correct series and will be recovered in future versions of the pipeline.
    \item There were 104 missing at least one of the six series and were thus incomplete.
    \item There were seven missing one or more channels (i.e., in-phase, opposed-phase, fat, water) from a single series.  These datasets may be recovered if the missing data are available for download from the UK Biobank.
    \item There were 49 acquisitions that failed for a variety of other reasons.  
\end{itemize}

\begin{table}[tbp]
\begin{center}
\caption{Total number of DICOM datasets by series in the current analysis along with the number of those (un)successfully processed.  Percent total is the number of DICOM datasets in a series divided by the total number of subjects (38,935).  Percent processed is the number of processed datasets divided by the number of DICOM datasets in that series.}
\begin{tabular}{ l r r r r r } \hline
Series & \multicolumn{2}{c}{Total} & \multicolumn{2}{c}{Processed}\\ \hline
Dixon & 38,929 & 99.98\% & 38,695 & 99.40\% \\
Pancreas 3D & 37,391 & 96.03\% & 37,385 & 99.98\% \\
Liver GRE & 10,130 & 26.02\% & 10,129 & 99.99\% \\
Liver IDEAL & 30,113 & 77.34\% & 30,113 & 100.00\% \\
Pancreas GRE & 31,017 & 79.66\% & 31,014 & 99.99\% \\ \hline
\end{tabular}
\label{tab:numbers}
\end{center}
\end{table}

The multiecho liver acquisition failed for only one subject, due to data corruption, and the conversion process did not fail for any subjects in the IDEAL liver acquisition.  The change in acquisition protocol after the pilot scanning period explains why 10,130 subjects have data from the gradient-echo protocol and 30,113 subjects have data from the IDEAL protocol (Table~\ref{tab:numbers}).  There were also 1,489 subjects who had data from both liver protocols, taking these duplicates into account means that a total of 38,754 single-slice liver series were acquired.  


The high-resolution 3D T1w pancreas acquisition failed for six subjects due to inconsistent acquisition parameters.  The multiecho pancreas acquisition failed for three subjects, also due to data corruption.  

The abdominal MRI scanning session takes approximately 10 minutes to perform but may not always successfully complete, for a variety of reasons such as lack of time, the subject cannot tolerate the session, failure to comply with breath hold instructions, etc. There were 31,019 (79.7\%) subjects who successfully completed the abdominal MRI scanning session with all four acquisitions, 6,343 (16.3\%) completed three of the four acquisitions, 1,413 (3.6\%) completed two of the four and 160 (0.4\%) completed only a single acquisition.

\subsection{Anomaly Detection}

\begin{table}[tbp]
\begin{center}
\caption{Number of anomalies detected by category in the Dixon acquisitions.}
\begin{tabular}{ l r } \hline
Anomaly Type & Count\\ \hline
Head/Neck in Field of View & 321 \\
Signal Dropout in the Chest & 145 \\
Signal Dropout in the Knee & 91 \\
Signal Dropout (other) & 12 \\ 
Series Shifted Left/Right & 3 \\
Other & 4 \\
More Than One Anomaly & 136 \\ \hline
\end{tabular}
\label{tab:anomalies}
\end{center}
\end{table}

We detected anomalies on 448 (1.3\%) of the Dixon scans.  The severity of these anomalies, and their impact on quantitative results, varied (Table~\ref{tab:anomalies}).  For example, artifacts in the chest appeared as small pockets of signal dropout or warped signal intensities in line with the sternum and may have had a small impact on the estimation of subcutaneous adipose tissue volume.  The presence of a substantial amount of the neck or head in the first Dixon series (accounting for the majority of anomalies detected) may have affected downstream analyses involving the thigh muscles.  Observing a single series that is shifted left-right with respect to the adjacent series could have had negative consequences on subsequent analyses such as organ segmentation.  

\subsection{Swap Detection}

\begin{figure}[tbp]
  \centering
  \includegraphics[width=\wf\textwidth]{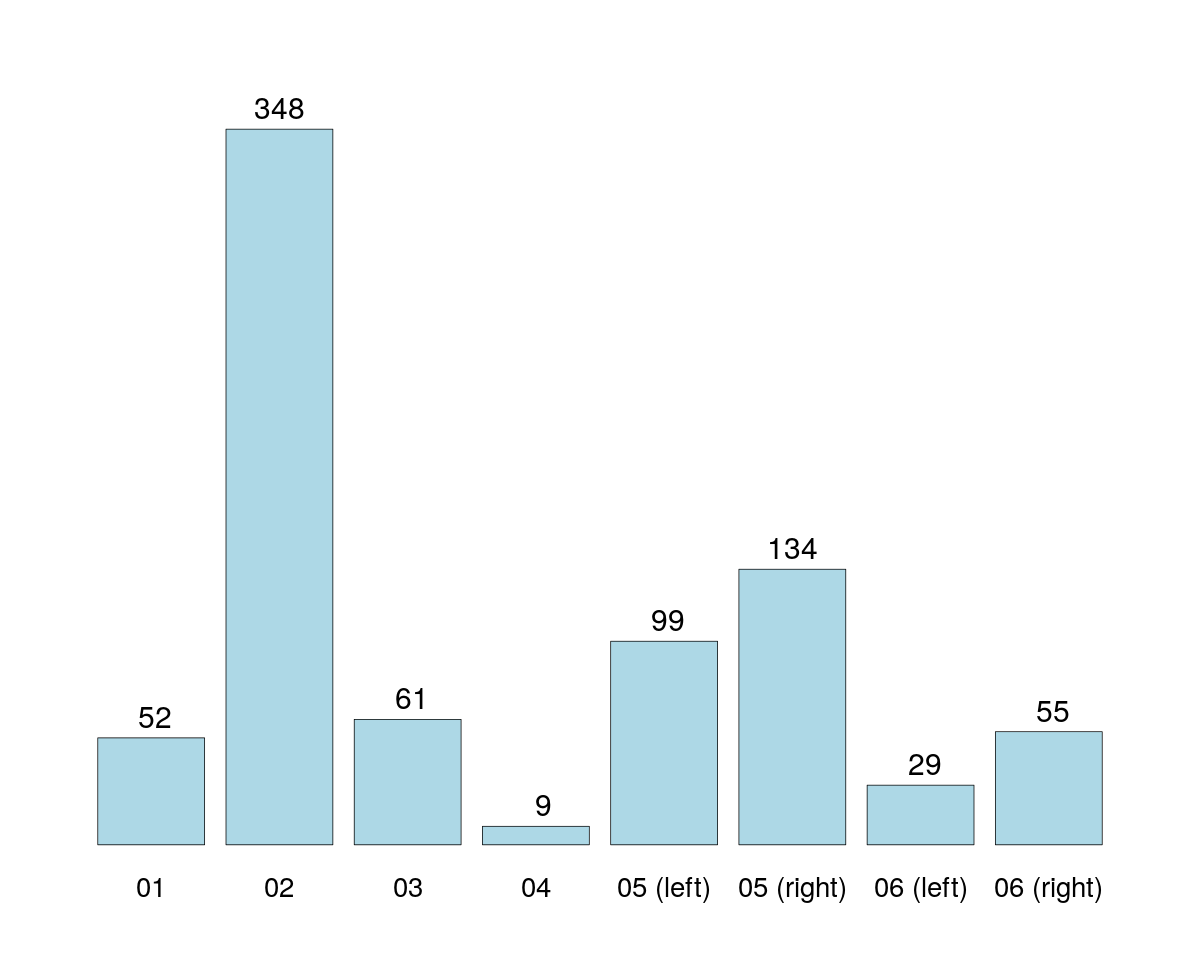}
  \caption{Total number of fat-water swaps by series, where the fifth and sixth series are split into left/right components.  The anatomical locations of each series roughly correspond to shoulders/neck (01), chest (02), abdomen (03), pelvis (04), upper thighs (05) and lower thighs (06).}
  \label{fig:swaps}
\end{figure}

Figure~\ref{fig:swaps} summarizes the total number of fat-water swaps detected in the successfully processed datasets.  Given that swaps may occur in each of the six series independently, a total number of $38,695 \times 8 = 390,560$ Dixon acquisitions were processed.  Of these, 787 (0.16\%) of them contained a fat-water swap, occurring in 690 (1.8\%) subjects.  The series with the most swaps was the second (348) which covers the chest, followed by the fifth series (233) which covers the upper thighs.

\subsection{Bone Joint Detection}

\begin{table}[tbp]
\begin{center}
\caption{Prediction error of the six bone joints as measured by average Euclidean distance (in millimeters) on 48 out-of-sample datasets.}
\begin{tabular}{ l r } \hline
Bone Joint & Prediction Error (mm)\\ \hline
Right Shoulder & 6.70 \\
Left Shoulder & 7.36 \\
Right Hip & 7.72 \\
Left Hip & 8.82 \\
Right Knee & 7.34 \\
Left Knee & 6.99 \\ \hline
\end{tabular}
\label{tab:bone_joints_error}
\end{center}
\end{table}

The average Euclidean distance between the test data and predicted bone joints is summarized in Table~\ref{tab:bone_joints_error}.  The prediction errors in Table~\ref{tab:bone_joints_error} were equivalent to an offset of 2--4 voxels on average.  The bone structures themselves were on the order of $20 \times 20 \times 14$~voxels in size, thus an average offset of 2--4 voxels is well within an acceptable level of tolerance.  

\begin{figure}[tbp]
  \centering
  \includegraphics[trim={0cm 2cm 18cm 0cm},clip,width=\wf\textwidth]{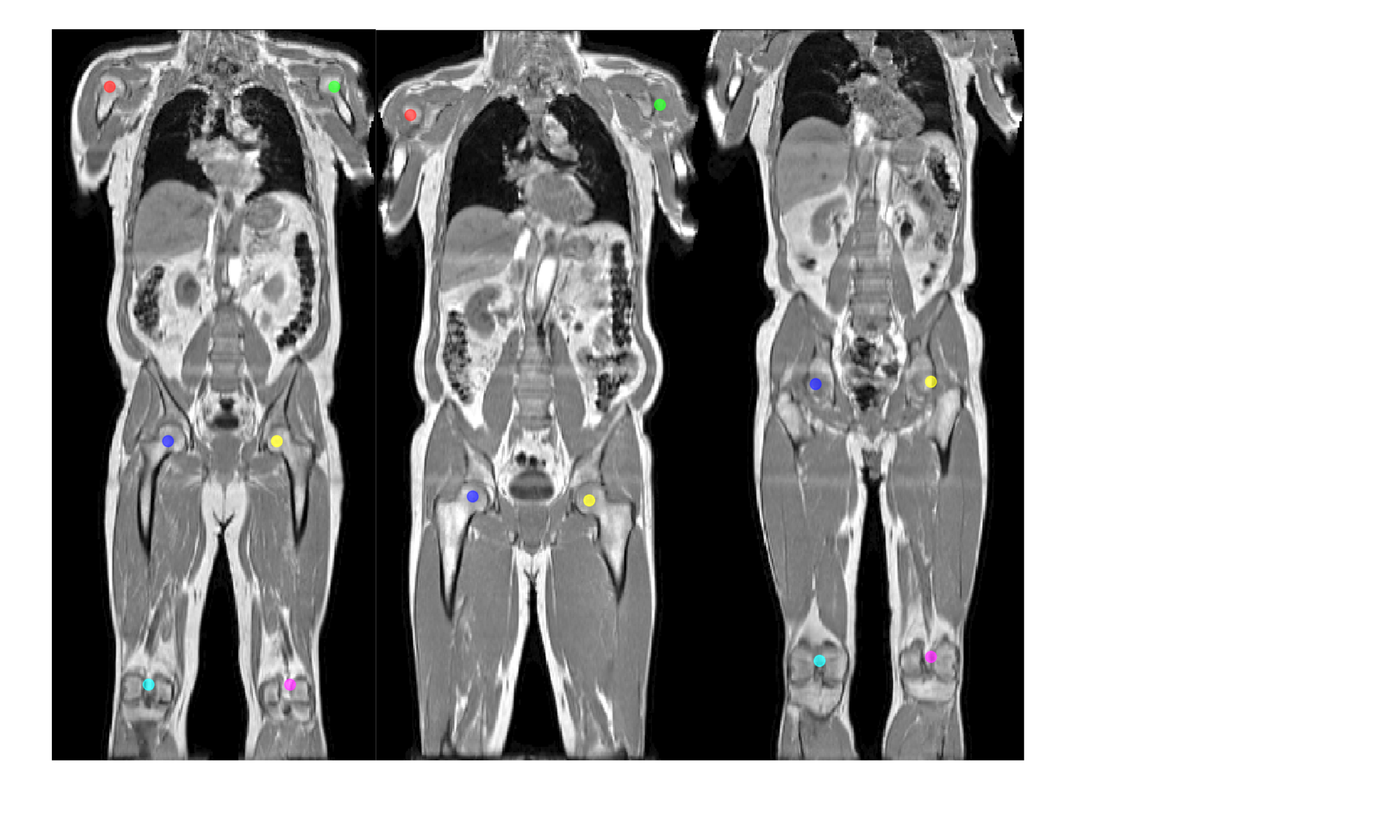}
  \caption{Predicted bone joints from a hybrid atlas-based registration and deep-learning algorithm.  The bone joints are labelled (from left-to-right, top-to-bottom): right shoulder (red), left shoulder (green), right hip (blue), left hip (yellow), right knee (turquoise), left knee (magenta).  A complete set of six bone joints are shown (left), along with missing knees (middle) and missing shoulders (right).}
  \label{fig:bone_joints}
\end{figure}

An example where the bone joints are displayed on a single coronal slice is provided in Figure~\ref{fig:bone_joints}.  A total number of 31,834 (96.0\%) of the acquisitions contained all six of the landmarks using the criterion that over half of the landmarks associated with the 35 atlases were present in the field of view after registration.  The presence of each individual bone joint is summarized in Table~\ref{tab:bone_joints}.  Given the limited coverage (1.1~meters) and small amount of variability in positioning the subject in the scanner, we expected to see a slight reduction in the number of predicted knee joints primarily due to the height of the subject.  

\begin{table}[tbp]
\begin{center}
\caption{Anatomical coverage of bone joints ($N = 38,504$).}
\begin{tabular}{ l r r r r} \hline
Location & \multicolumn{2}{c}{Right} & \multicolumn{2}{c}{Left} \\ \hline
Shoulder & 38,263 & 99.4\% & 38,248 & 99.3\% \\
Hip & 38,504 & 100.0\% & 38,504 & 100.0\% \\
Knee & 37,379 & 97.1\% & 37,372 & 97.1\% \\ \hline
\end{tabular}
\label{tab:bone_joints}
\end{center}
\end{table}

Approximately 3\% of the imaging cohort scanned had knee joints outside the fixed coverage of the Dixon protocol, while only 0.7\% of shoulder joints were not available.  A total of 1,564 (4.1\%) subjects had at least one missing bone joint, 305 subjects had at least one missing shoulder joint (63\% missing both shoulder joints and 37\% missing only one) and 1,259 subjects had at least one missing knee joint (79\% missing both knee joints and 21\% missing only one).  No subjects were found to be missing both their shoulders and knees.  

\subsection{Anatomical Coverage of the Single-Slice Acquisition}

\begin{figure}[tbp]
  \centering
  \includegraphics[width=\wf\textwidth]{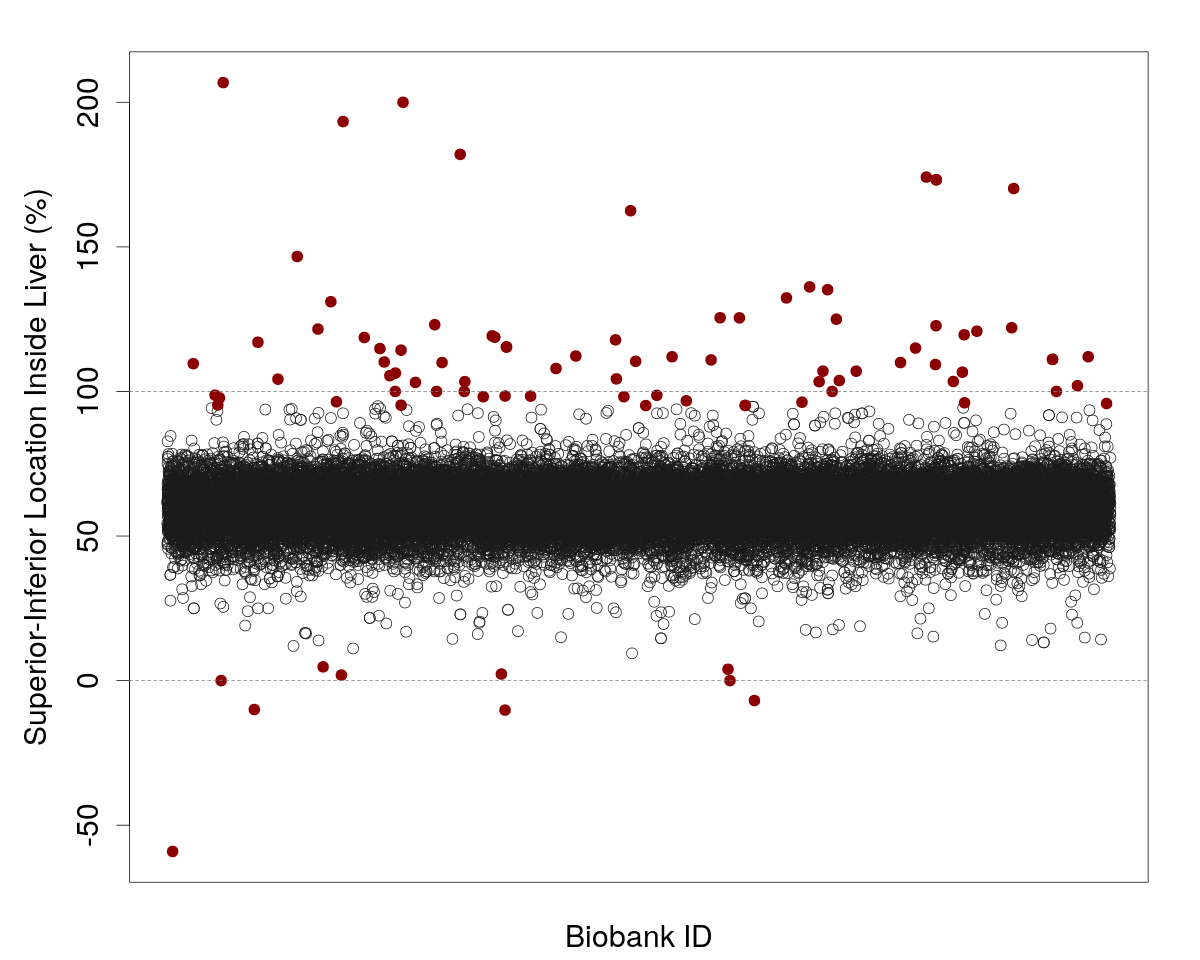}
  \caption{Superior-Inferior location of the single-slice multiecho liver acquisition with respect to a 3D segmentation of the liver ($N = 33,284$).  Each point represents a subject in the UK Biobank imaging cohort ($x$-axis).  The $y$-axis is the location relative to the three-dimensional segmentation in the Dixon space, where the upper and lower axial slices of the 3D liver segmentation represent 0 and 100\% for that subject (dashed lines).  Slice locations less than 5\% or greater than 95\% of the 3D liver segmentation are highlighted.}
  \label{fig:single_slice_liver_placement}
\end{figure}

Figure~\ref{fig:single_slice_liver_placement} displays the slice locations, as a percentage of the height of the 3D segmentation of the liver (cf. Section~\ref{sec:slice_location}).  We used neural-network based 3D segmentations of the liver to evaluate the slice placement of the multiecho sequences \citep{Liu2020systematic}.  Slice locations that were less than 0\% or greater than 100\% indicated the single-slice multiecho acquisition did not cover the subject's liver.  Given the potential for slight misalignment due to differences in each breath hold by a subject, slice locations below the bottom 5\% or above the top 95\% of the liver were identified as unreliable.  There were 91 (0.27\%) of the acquisitions outside these thresholds.  

\begin{figure}[tbp]
  \centering
  \includegraphics[width=\wf\textwidth]{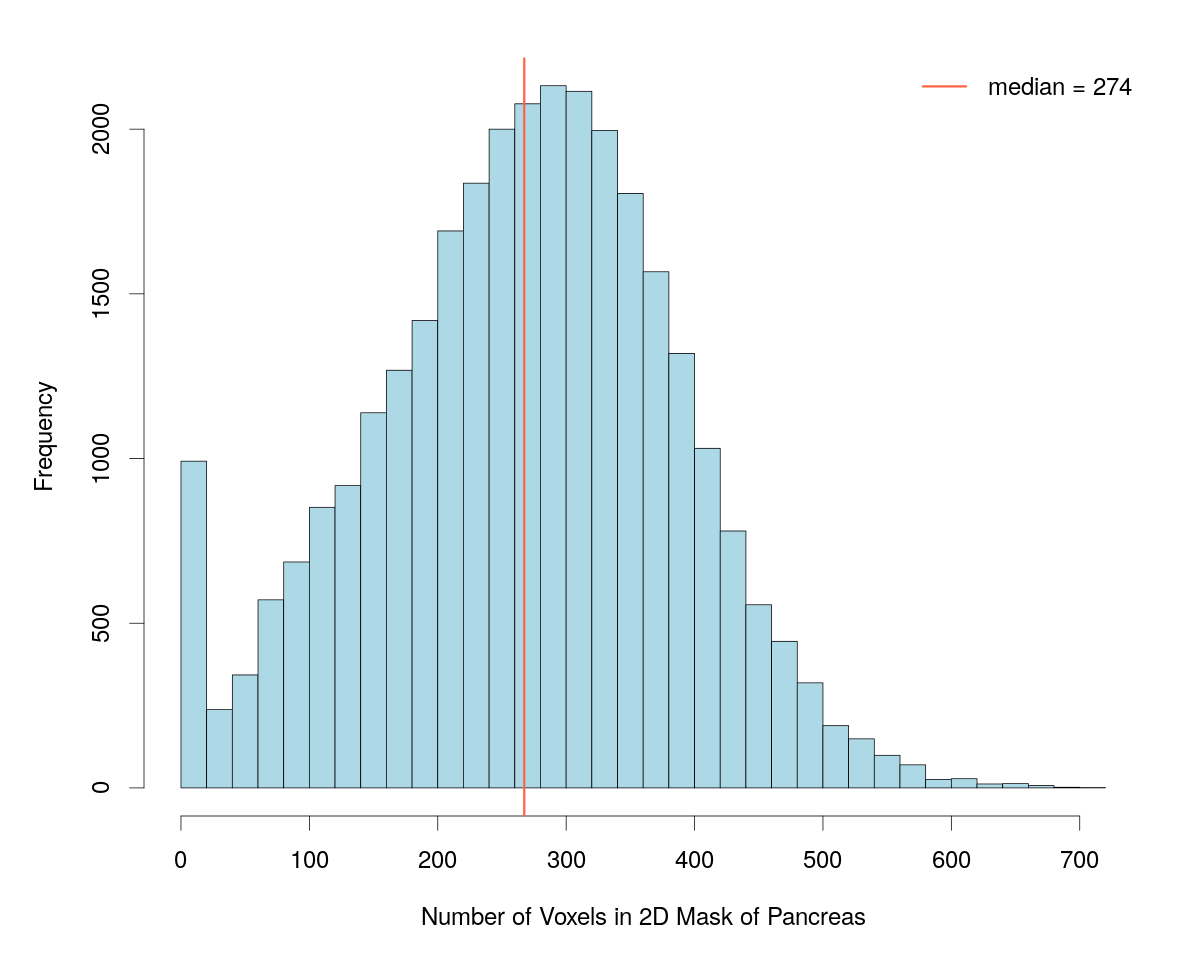}
  \caption{The number of voxels contained in the 2D mask of the pancreas for the single-slice multiecho acquisition ($N = 30,691$).  The median number of voxels across all available subjects is indicated by the vertical red line.}
  \label{fig:single_slice_pancreas_n}
\end{figure}

The median number of voxels in a 2D~mask of the pancreas over the available subjects was 274, which is approximately $10.3~\textrm{ml}$ in volume, for the 2D masks of the pancreas.  Given the typical volume of an adult pancreas is in the range of 71--83~ml \citep{desouza20pancreas} the volumes in the single-slice multiecho acquisitions are on the order of 10-15\% of the total pancreas volume.  There were a substantial number of 2D masks between 0--25 voxels (Figure~\ref{fig:single_slice_pancreas_n}) in size, and a total of 731 (2.4\%) contained zero voxels implying that the placement of the single-slice multiecho acquisition did not intersect with the pancreas.  

\subsection{Proton Density Fat Fraction}

\begin{figure}[tbp]
  \centering
  \includegraphics[width=0.49\linewidth]{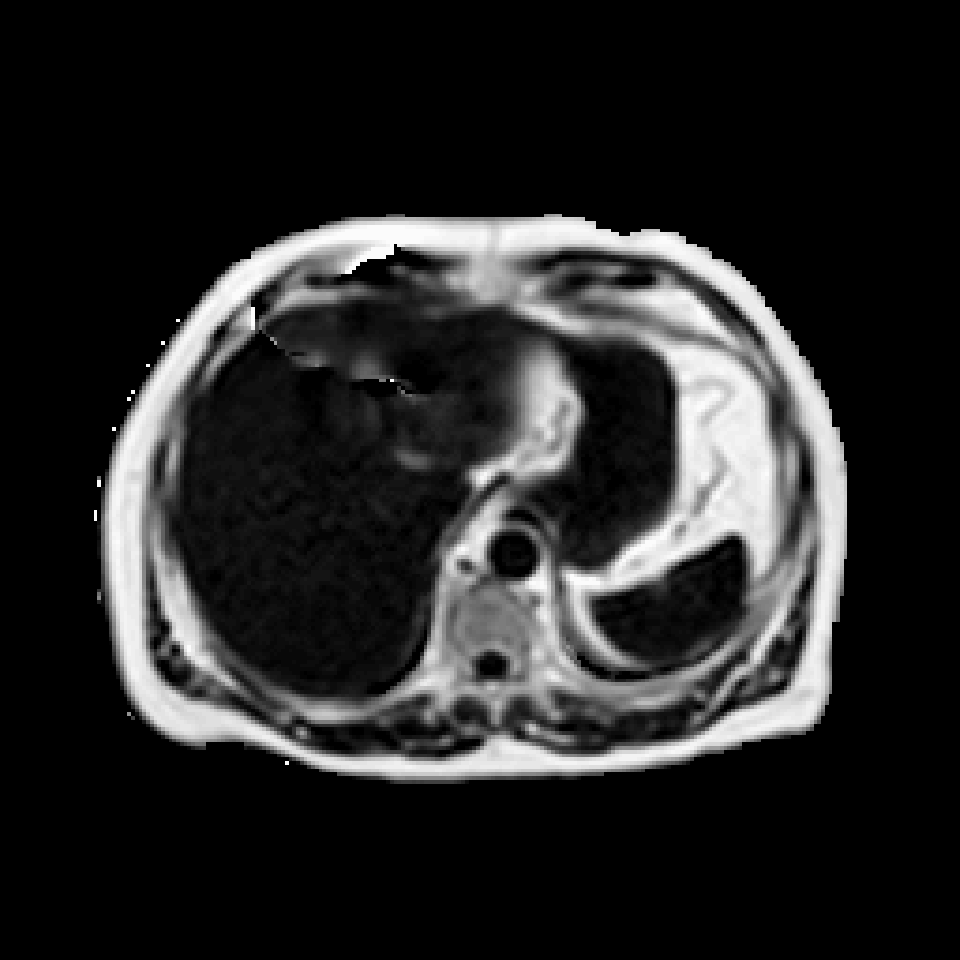}\includegraphics[width=0.49\linewidth]{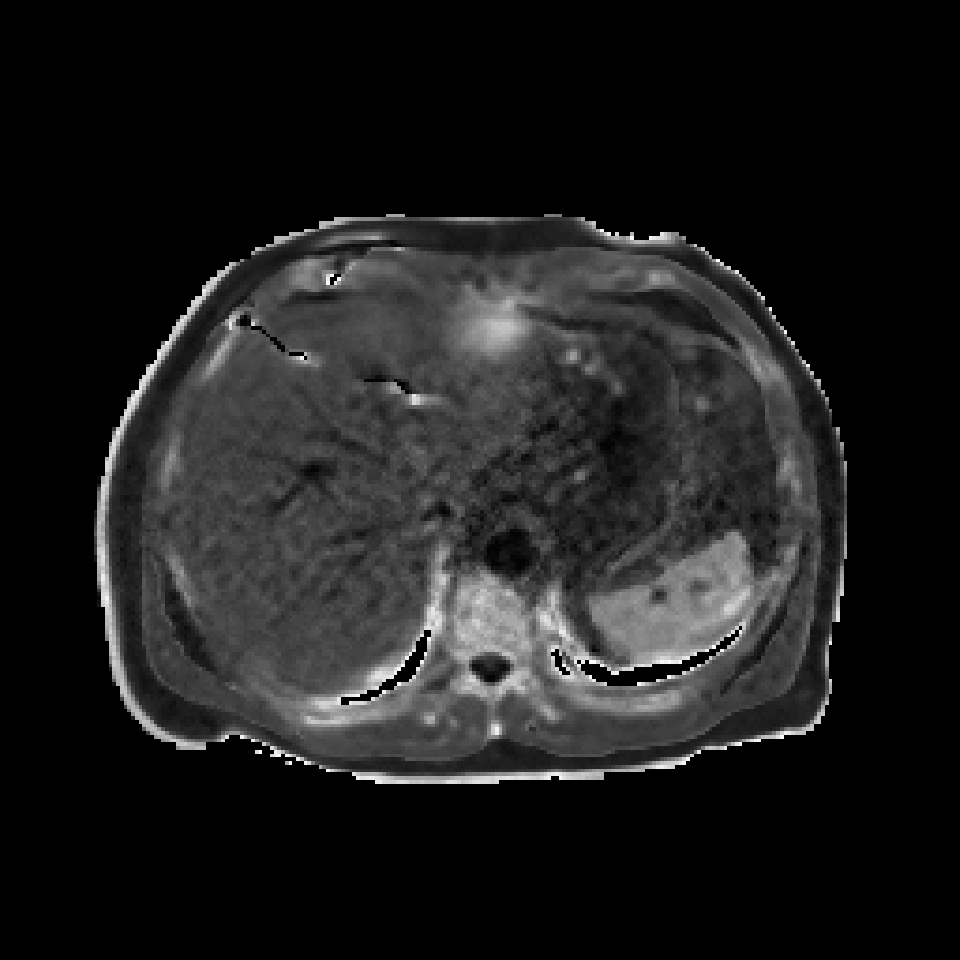}
\caption{Parameter estimates from the quantification of a single-slice IDEAL acquisition centered on the liver (from left-to-right with min/max units): PDFF ($0-100$\%) and R2* ($0-150~\mathrm{s}^{-1}$).  For visualization purposes all voxels outside the body have been suppressed.}
\label{fig:presco}
\end{figure}

Estimation of the PDFF and transverse relaxivity (R2*) was performed on all available single-slice acquisitions for both the liver and pancreas.  Figure~\ref{fig:presco} summarizes these parameters by the PRESCO algorithm using both the magnitude and phase information from a single-slice IDEAL acquisition.  The PDFF estimates span the full range of values ($0-100$\%) since the PRESCO algorithm utilizes the complex-valued MR data.  Thus, low PDFF values, in abdominal organs such as the liver and spleen, were present in addition to high PDFF values associated with subcutaneous and visceral adipose tissue.  The estimated R2* values differentiated tissue types (e.g., vessels) in the liver.  

\begin{figure}[tbp]
  \centering
  \includegraphics[width=\wf\textwidth]{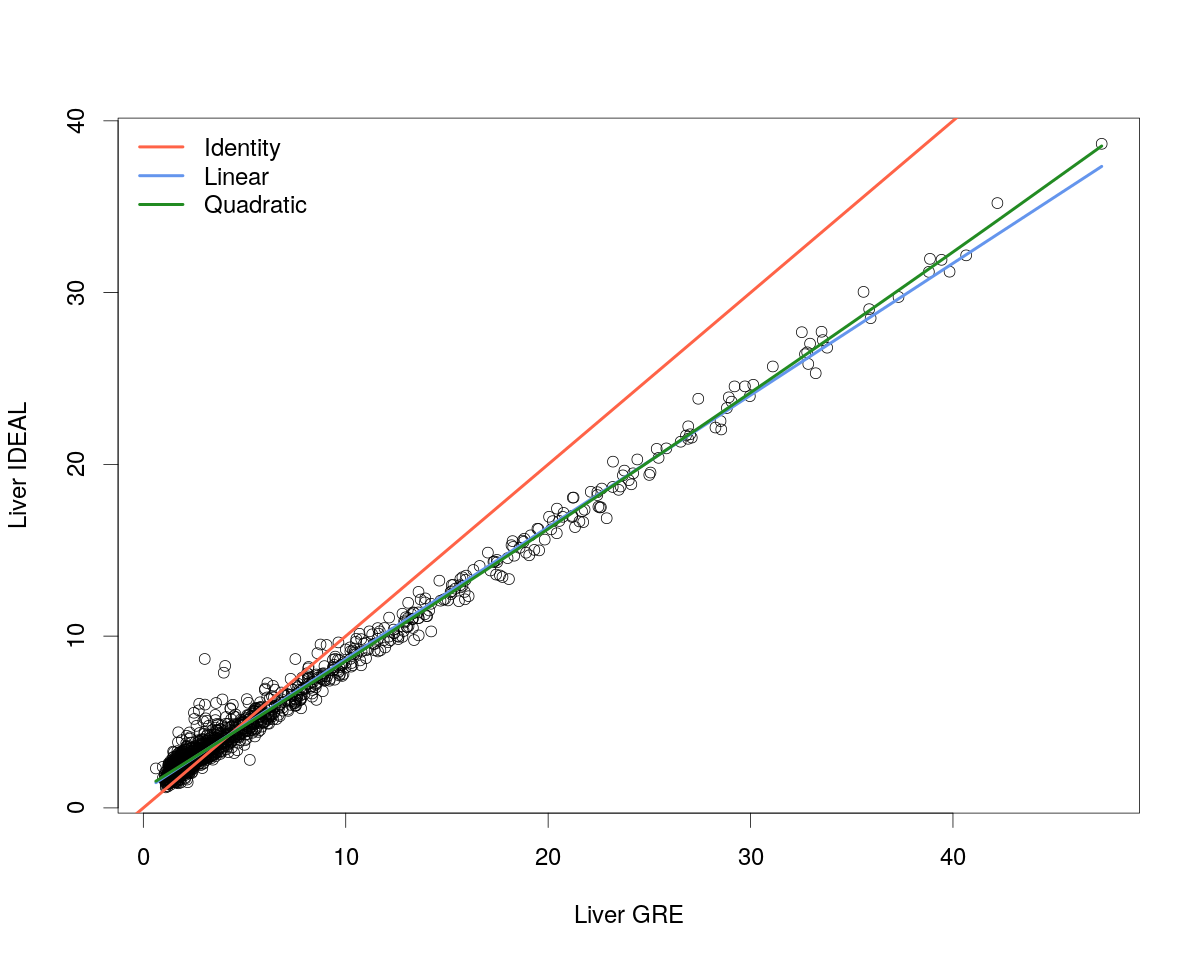}
  \caption{Median PDFF values (in percent) estimated using the PRESCO algorithm from two distinct acquisitions, IDEAL and GRE multiecho, applied to the same group of subjects ($N=1,463$).  Regression lines for the identity, linear and quadratic models are displayed.}
  \label{fig:ideal_vs_multiecho}
\end{figure}

When processing the single-slice multiecho liver acquisitions nine (0.02\%) were found to have deviated from the acquisition protocol in that they were not acquired in the axial (transverse) plane and the liver was minimally covered by the resulting image.  These datasets were not suitable for analysis.  With respect to the single-slice multiecho pancreas acquisitions 566 (1.8\%) deviated from the acquisition protocol in that they were not acquired in the axial (transverse) plane.  One acquisition contained a single echo time (out of the 10 normally acquired) with corrupted data.  All of these datasets have the potential to be processed by the PRESCO algorithm and will be recovered in future versions of the pipeline.

When the acquisition protocol was updated to the IDEAL sequence from the multiecho sequence, after the first approximately 10,000 subjects were scanned (cf. Table~\ref{tab:numbers}) a group of subjects had both protocols acquired.  Figure~\ref{fig:ideal_vs_multiecho} compares the median PDFF values (in percent) across the liver for all available subjects between the two acquisition sequences.  To ensure consistency, the binary masks that define liver tissue \citep{Liu2020systematic} were projected into the other acquisition space (i.e., IDEAL into GRE, GRE into IDEAL) and the intersection was applied to produce the same anatomical coverage of the liver in both acquisitions. The IDEAL PDFF estimates were higher, compared with the multiecho sequence, and this relationship reversed around 4.3\% where the IDEAL PDFF estimates were lower compared with the multiecho sequence.  Fitting a simple linear regression model between these two PDFF estimates yielded
\begin{equation}
    \mathrm{PDFF}_\mathrm{IDEAL} = 1.0 + 0.7678 \cdot \mathrm{PDFF}_\mathrm{GRE}, 
\end{equation}
adding a quadratic term to this model yields
\begin{equation}
    \mathrm{PDFF}_\mathrm{IDEAL} = 1.1 + 0.7326 \cdot \mathrm{PDFF}_\mathrm{GRE} + 0.0012 \cdot \mathrm{PDFF}_\mathrm{GRE}^2, 
\end{equation}
where the adjusted $R^2$ values for the models were 0.9873 and 0.9876, respectively, and the quadratic term was statistically significant when comparing the two models ($F_{1,1460}=36.974$, $p<0.001$).




\section{Discussion}

The investigation of body composition in abdominal organs will be greatly enhanced through the image-derived phenotypes produced from the abdominal MRI acquisitions in the UK Biobank.  The large number of datasets requires automation to be utilized at every stage of the processing and quality control procedures.  An image analysis pipeline that accommodates three of the acquisition types from the abdominal MRI protocol (neck-to-knee Dixon, high-resolution 3D T1-weighted and single-slice multiecho) has been introduced here that includes several automated steps.  Quality control procedures have been implemented in the pipeline that capture and quantify a wide variety of features, providing actionable feedback to the user so that datasets may be discarded in downstream analyses or flagged up for visual inspection.  

There are several aspects of the image analysis pipeline and quality control procedures that may be improved.  The Dixon acquisition protocol is challenging, given multiple series are involved that produce the neck-to-knee coverage.  From processing the first 38,971 it is apparent that extra series (more than six) are a persistent issue and may arise from a variety of reasons.  Common situations uncovered here include
\begin{itemize}
    \item halting the acquisition, re-positioning the subject and restarting the acquisition,
    \item acquiring a seventh series that covers the lower legs, or
    \item acquiring a seventh series that duplicates one of the previous six stations.
\end{itemize}
Identifying these situations is straightforward by simply counting the number of Dixon series, but correctly processing them into a consistent neck-to-knee volume automatically will require a combination of algorithms and decision rules in order to characterize the additional series, select the six correct series and discard the irrelevant series.  Strategies for this level of automation are currently being investigated.  

Swap detection is currently optimized and applied to the anatomical locations associated with the six series in the Dixon acquisition protocol.  The desire for parsimony and generalization leads one to consider a single model that is applicable to all series independent of anatomical coverage.  Combining several steps in the assembly process (e.g., blending, bias-field correction, swap detection/correction) is also of interest both to improve overall data consistency and accelerate the pipeline.  

Anomaly detection is currently implemented as a method to flag suspicious Dixon neck-to-knee datasets and indicate visual inspection is required to make a decision on including the data in subsequent analyses.  Obvious next steps would be to refine and/or expand the categories covered by the anomaly detection methodology and implement an automated correction technique for as many of the anomaly categories as possible.  Given the relatively small numbers of anomalies detected these efforts will need to be prioritized appropriately and are currently under review for development.  

Estimating the PDFF has been implemented for single-slice multiecho acquisitions in the axial (transverse) plane.  This is due to reordering the multidimensional arrays into simple two-dimensional matrices to take advantage of efficient computational techniques in higher-level languages.  Refactoring the code so that it accommodates single-slice multiecho data of any orientation (i.e., axial, sagittal, coronal) or multi-slice data is currently being scheduled for development.  

The image analysis pipeline, including quality control methods, presented here produces standardized data and parameter estimates for subsequent analyses \citep{Liu2020systematic}.  We look forward to applying the pipeline to the remaining subjects from the UK Biobank imaging cohort (100,000 subjects in total) and the anticipated repeat scans.

\section{Acknowledgements}

This research has been conducted using the UK Biobank Resource under Application Number `44584' and was funded by Calico Life Sciences LLC.  We would like to thank Dr Mark Bydder for making his software available in the public domain and valuable input regarding estimation of the proton density fat fraction.  

\bibliographystyle{model2-names.bst}
\biboptions{authoryear}
\bibliography{main}

\begin{thebibliography}{32}
\expandafter\ifx\csname natexlab\endcsname\relax\def\natexlab#1{#1}\fi
\providecommand{\url}[1]{\texttt{#1}}
\providecommand{\href}[2]{#2}
\providecommand{\path}[1]{#1}
\providecommand{\DOIprefix}{doi:}
\providecommand{\ArXivprefix}{arXiv:}
\providecommand{\URLprefix}{URL: }
\providecommand{\Pubmedprefix}{pmid:}
\providecommand{\doi}[1]{\href{http://dx.doi.org/#1}{\path{#1}}}
\providecommand{\Pubmed}[1]{\href{pmid:#1}{\path{#1}}}
\providecommand{\bibinfo}[2]{#2}
\ifx\xfnm\relax \def\xfnm[#1]{\unskip,\space#1}\fi
\bibitem[{Alfaro-Almagro et~al.(2018)Alfaro-Almagro, Jenkinson, Bangerter,
  Andersson, Griffanti, Douaud, Sotiropoulos, Jbabdi, Hernandez-Fernandez,
  Vallee, Vidaurre, Webster, McCarthy, Rorden, Daducci, Alexander, Zhang,
  Dragonu, Matthews, Miller and Smith}]{alfaro2018image}
\bibinfo{author}{Alfaro-Almagro, F.}, \bibinfo{author}{Jenkinson, M.},
  \bibinfo{author}{Bangerter, N.K.}, \bibinfo{author}{Andersson, J.L.R.},
  \bibinfo{author}{Griffanti, L.}, \bibinfo{author}{Douaud, G.},
  \bibinfo{author}{Sotiropoulos, S.N.}, \bibinfo{author}{Jbabdi, S.},
  \bibinfo{author}{Hernandez-Fernandez, M.}, \bibinfo{author}{Vallee, E.},
  \bibinfo{author}{Vidaurre, D.}, \bibinfo{author}{Webster, M.},
  \bibinfo{author}{McCarthy, P.}, \bibinfo{author}{Rorden, C.},
  \bibinfo{author}{Daducci, A.}, \bibinfo{author}{Alexander, D.C.},
  \bibinfo{author}{Zhang, H.}, \bibinfo{author}{Dragonu, I.},
  \bibinfo{author}{Matthews, P.M.}, \bibinfo{author}{Miller, K.L.},
  \bibinfo{author}{Smith, S.M.}, \bibinfo{year}{2018}.
\newblock \bibinfo{title}{Image processing and quality control for the first
  10,000 brain imaging datasets from {UK} {Biobank}}.
\newblock \bibinfo{journal}{NeuroImage} \bibinfo{volume}{166},
  \bibinfo{pages}{400--424}.
\newblock \DOIprefix\doi{10.1016/j.neuroimage.2017.10.034}.
\bibitem[{Aljabar et~al.(2009)Aljabar, Heckemann, Hammers, Hajnal and
  Rueckert}]{aljabar2009multi}
\bibinfo{author}{Aljabar, P.}, \bibinfo{author}{Heckemann, R.A.},
  \bibinfo{author}{Hammers, A.}, \bibinfo{author}{Hajnal, J.V.},
  \bibinfo{author}{Rueckert, D.}, \bibinfo{year}{2009}.
\newblock \bibinfo{title}{Multi-atlas based segmentation of brain images: Atlas
  selection and its effect on accuracy}.
\newblock \bibinfo{journal}{NeuroImage} \bibinfo{volume}{46},
  \bibinfo{pages}{726--738}.
\bibitem[{Bagur et~al.(2019)Bagur, Hutton, Irving, Gyngell, Robson and
  Brady}]{triay2019magnitude}
\bibinfo{author}{Bagur, A.T.}, \bibinfo{author}{Hutton, C.},
  \bibinfo{author}{Irving, B.}, \bibinfo{author}{Gyngell, M.L.},
  \bibinfo{author}{Robson, M.D.}, \bibinfo{author}{Brady, M.},
  \bibinfo{year}{2019}.
\newblock \bibinfo{title}{Magnitude-intrinsic water-fat ambiguity can be
  resolved with multipeak fat modeling and a multipoint search method}.
\newblock \bibinfo{journal}{Magnetic Resonance in Medicine}
  \bibinfo{volume}{82}, \bibinfo{pages}{460--475}.
\bibitem[{Basty et~al.(2020)Basty, Liu, Cule, Thomas, Bell and
  Whitcher}]{basty2020automated}
\bibinfo{author}{Basty, N.}, \bibinfo{author}{Liu, Y.}, \bibinfo{author}{Cule,
  M.}, \bibinfo{author}{Thomas, E.L.}, \bibinfo{author}{Bell, J.D.},
  \bibinfo{author}{Whitcher, B.}, \bibinfo{year}{2020}.
\newblock \bibinfo{title}{Automated measurement of pancreatic fat and iron
  concentration using multi-echo and {T1}-weighted {MRI} data}, in:
  \bibinfo{booktitle}{2020 IEEE 17th International Symposium on Biomedical
  Imaging (ISBI)}, \bibinfo{organization}{IEEE}. pp. \bibinfo{pages}{345--348}.
\bibitem[{Borga et~al.(2018)Borga, West, Bell, Harvey, Romu, Heymsfield and
  {Dahlqvist Leinhard}}]{borga2018advanced}
\bibinfo{author}{Borga, M.}, \bibinfo{author}{West, J.}, \bibinfo{author}{Bell,
  J.D.}, \bibinfo{author}{Harvey, N.C.}, \bibinfo{author}{Romu, T.},
  \bibinfo{author}{Heymsfield, S.B.}, \bibinfo{author}{{Dahlqvist Leinhard},
  O.}, \bibinfo{year}{2018}.
\newblock \bibinfo{title}{Advanced body composition assessment: From body mass
  index to body composition profiling}.
\newblock \bibinfo{journal}{Journal of Investigative Medicine}
  \bibinfo{volume}{66}, \bibinfo{pages}{1--9}.
\bibitem[{Bydder et~al.(2020)Bydder, Ghodrati, Gao, Robson, Yang and
  Hu}]{bydder2020constraints}
\bibinfo{author}{Bydder, M.}, \bibinfo{author}{Ghodrati, V.},
  \bibinfo{author}{Gao, Y.}, \bibinfo{author}{Robson, M.D.},
  \bibinfo{author}{Yang, Y.}, \bibinfo{author}{Hu, P.}, \bibinfo{year}{2020}.
\newblock \bibinfo{title}{Constraints in estimating the proton density fat
  fraction}.
\newblock \bibinfo{journal}{Magnetic Resonance Imaging} \bibinfo{volume}{66},
  \bibinfo{pages}{1--8}.
\bibitem[{Bydder et~al.(2008)Bydder, Yokoo, Hamilton, Middleton, Chavez,
  Schwimmer, Lavine and Sirlin}]{bydder2008relaxation}
\bibinfo{author}{Bydder, M.}, \bibinfo{author}{Yokoo, T.},
  \bibinfo{author}{Hamilton, G.}, \bibinfo{author}{Middleton, M.S.},
  \bibinfo{author}{Chavez, A.D.}, \bibinfo{author}{Schwimmer, J.B.},
  \bibinfo{author}{Lavine, J.E.}, \bibinfo{author}{Sirlin, C.B.},
  \bibinfo{year}{2008}.
\newblock \bibinfo{title}{Relaxation effects in the quantification of fat using
  gradient echo imaging}.
\newblock \bibinfo{journal}{Magnetic Resonance Imaging} \bibinfo{volume}{26},
  \bibinfo{pages}{347--359}.
\bibitem[{DeSouza et~al.(2018)DeSouza, Singh, Singh, Yoon, Murphy, Plank and
  Petrov}]{desouza20pancreas}
\bibinfo{author}{DeSouza, S.V.}, \bibinfo{author}{Singh, R.G.},
  \bibinfo{author}{Singh, R.G.}, \bibinfo{author}{Yoon, H.D.},
  \bibinfo{author}{Murphy, R.}, \bibinfo{author}{Plank, L.D.},
  \bibinfo{author}{Petrov, M.S.}, \bibinfo{year}{2018}.
\newblock \bibinfo{title}{Pancreas volume in health and disease: A systematic
  review and meta-analysis}.
\newblock \bibinfo{journal}{Expert Review of Gastroenterology \& Hepatology}
  \bibinfo{volume}{12}, \bibinfo{pages}{757--766}.
\newblock \DOIprefix\doi{10.1080/17474124.2018.1496015}.
\bibitem[{Ji et~al.(2019)Ji, Yiorkas, Frau, Mook-Kanamori, Staiger, Thomas,
  Atabaki-Pasdar, Campbell, Tyrrell, Jones, Beaumont, Wood, Tuke, Ruth,
  Mahajan, Murray, Freath, Weedon, Hattersley, Hayward, Machann, H{\"a}ring,
  Franks, {de Mutsert}, Pearson, Stefan, Frayling, Allebrandt, Bell, Blakemore
  and Yaghootkar}]{ji2019genome}
\bibinfo{author}{Ji, Y.}, \bibinfo{author}{Yiorkas, A.M.},
  \bibinfo{author}{Frau, F.}, \bibinfo{author}{Mook-Kanamori, D.},
  \bibinfo{author}{Staiger, H.}, \bibinfo{author}{Thomas, E.L.},
  \bibinfo{author}{Atabaki-Pasdar, N.}, \bibinfo{author}{Campbell, A.},
  \bibinfo{author}{Tyrrell, J.}, \bibinfo{author}{Jones, S.E.},
  \bibinfo{author}{Beaumont, R.N.}, \bibinfo{author}{Wood, A.R.},
  \bibinfo{author}{Tuke, M.A.}, \bibinfo{author}{Ruth, K.S.},
  \bibinfo{author}{Mahajan, A.}, \bibinfo{author}{Murray, A.},
  \bibinfo{author}{Freath, R.M.}, \bibinfo{author}{Weedon, M.N.},
  \bibinfo{author}{Hattersley, A.T.}, \bibinfo{author}{Hayward, C.},
  \bibinfo{author}{Machann, J.}, \bibinfo{author}{H{\"a}ring, H.U.},
  \bibinfo{author}{Franks, P.}, \bibinfo{author}{{de Mutsert}, R.},
  \bibinfo{author}{Pearson, E.}, \bibinfo{author}{Stefan, N.},
  \bibinfo{author}{Frayling, T.M.}, \bibinfo{author}{Allebrandt, K.V.},
  \bibinfo{author}{Bell, J.D.}, \bibinfo{author}{Blakemore, A.I.},
  \bibinfo{author}{Yaghootkar, H.}, \bibinfo{year}{2019}.
\newblock \bibinfo{title}{Genome-wide and abdominal {MRI} data provide evidence
  that a genetically determined favorable adiposity phenotype is characterized
  by lower ectopic liver fat and lower risk of type 2 diabetes, heart disease,
  and hypertension}.
\newblock \bibinfo{journal}{Diabetes} \bibinfo{volume}{68},
  \bibinfo{pages}{207--219}.
\newblock \DOIprefix\doi{10.2337/db18-0708}.
\bibitem[{Langner et~al.(2020a)Langner, Ahlstr{\"o}m and
  Kullberg}]{langner2020large}
\bibinfo{author}{Langner, T.}, \bibinfo{author}{Ahlstr{\"o}m, H.},
  \bibinfo{author}{Kullberg, J.}, \bibinfo{year}{2020}a.
\newblock \bibinfo{title}{Large-scale biometry with interpretable neural
  network regression on {UK} {Biobank} body {MRI}}.
\newblock \bibinfo{note}{{arXiv}:2002.0682}.
\bibitem[{Langner et~al.(2020b)Langner, Wikstr{\"o}m, Bjerner, Ahlstr{\"o}m and
  Kullberg}]{langner2019identifying}
\bibinfo{author}{Langner, T.}, \bibinfo{author}{Wikstr{\"o}m, J.},
  \bibinfo{author}{Bjerner, T.}, \bibinfo{author}{Ahlstr{\"o}m, H.},
  \bibinfo{author}{Kullberg, J.}, \bibinfo{year}{2020}b.
\newblock \bibinfo{title}{Identifying morphological indicators of aging with
  neural networks on large-scale whole-body {MRI}}.
\newblock \bibinfo{journal}{IEEE Transactions on Medical Imaging}
  \bibinfo{volume}{30}, \bibinfo{pages}{1430--1437}.
\newblock \DOIprefix\doi{10.1109/TMI.2019.2950092}.
\bibitem[{Linge et~al.(2018)Linge, Borga, West, Tuthill, Miller, Dumitriu,
  Thomas, Romu, Tun{\'o}n, Bell and {Dahlqvist Leinhard}}]{linge2018body}
\bibinfo{author}{Linge, J.}, \bibinfo{author}{Borga, M.},
  \bibinfo{author}{West, J.}, \bibinfo{author}{Tuthill, T.},
  \bibinfo{author}{Miller, M.R.}, \bibinfo{author}{Dumitriu, A.},
  \bibinfo{author}{Thomas, E.L.}, \bibinfo{author}{Romu, T.},
  \bibinfo{author}{Tun{\'o}n, P.}, \bibinfo{author}{Bell, J.D.},
  \bibinfo{author}{{Dahlqvist Leinhard}, O.}, \bibinfo{year}{2018}.
\newblock \bibinfo{title}{Body composition profiling in the {UK} {Biobank}
  imaging study}.
\newblock \bibinfo{journal}{Obesity} \bibinfo{volume}{26},
  \bibinfo{pages}{1785--1795}.
\bibitem[{Linge et~al.(2019)Linge, Whitcher, Borga and {Dahlqvist
  Leinhard}}]{linge2019sub}
\bibinfo{author}{Linge, J.}, \bibinfo{author}{Whitcher, B.},
  \bibinfo{author}{Borga, M.}, \bibinfo{author}{{Dahlqvist Leinhard}, O.},
  \bibinfo{year}{2019}.
\newblock \bibinfo{title}{Sub-phenotyping metabolic disorders using body
  composition: An individualized, nonparametric approach utilizing large data
  sets}.
\newblock \bibinfo{journal}{Obesity} \bibinfo{volume}{27},
  \bibinfo{pages}{1190--1199}.
\bibitem[{Littlejohns et~al.(2020)Littlejohns, Holliday, Gibson, Garratt,
  Oesingmann, Alfaro-Almagro, Bell, Boultwood, Collins, Conroy, Crabtree,
  Doherty, Frangi, Harvey, Leeson, Miller, Neubauer, Petersen, Sellors, Sheard,
  Smith, Sudlow, Matthews and Allen}]{littlejohns2020biobank}
\bibinfo{author}{Littlejohns, T.J.}, \bibinfo{author}{Holliday, J.},
  \bibinfo{author}{Gibson, L.M.}, \bibinfo{author}{Garratt, S.},
  \bibinfo{author}{Oesingmann, N.}, \bibinfo{author}{Alfaro-Almagro, F.},
  \bibinfo{author}{Bell, J.D.}, \bibinfo{author}{Boultwood, C.},
  \bibinfo{author}{Collins, R.}, \bibinfo{author}{Conroy, M.C.},
  \bibinfo{author}{Crabtree, N.}, \bibinfo{author}{Doherty, N.},
  \bibinfo{author}{Frangi, A.F.}, \bibinfo{author}{Harvey, N.C.},
  \bibinfo{author}{Leeson, P.}, \bibinfo{author}{Miller, K.L.},
  \bibinfo{author}{Neubauer, S.}, \bibinfo{author}{Petersen, S.E.},
  \bibinfo{author}{Sellors, J.}, \bibinfo{author}{Sheard, S.},
  \bibinfo{author}{Smith, S.M.}, \bibinfo{author}{Sudlow, C.L.M.},
  \bibinfo{author}{Matthews, P.M.}, \bibinfo{author}{Allen, N.E.},
  \bibinfo{year}{2020}.
\newblock \bibinfo{title}{The {UK} {Biobank} imaging enhancement of 100,000
  participants: Rationale, data collection, management and future directions}.
\newblock \bibinfo{journal}{Nature Communications}
  \DOIprefix\doi{10.1038/s41467-020-15948-9}.
\bibitem[{Liu et~al.(2020)Liu, Basty, Whitcher, Bell, van Bruggen, Thomas and
  Cule}]{Liu2020systematic}
\bibinfo{author}{Liu, Y.}, \bibinfo{author}{Basty, N.},
  \bibinfo{author}{Whitcher, B.}, \bibinfo{author}{Bell, J.},
  \bibinfo{author}{van Bruggen, N.}, \bibinfo{author}{Thomas, E.L.},
  \bibinfo{author}{Cule, M.}, \bibinfo{year}{2020}.
\newblock \bibinfo{title}{Systematic quantification of health parameters from
  {UK} {B}iobank abdominal {MRI} using deep learning}.
\newblock \bibinfo{journal}{bioRxiv} \DOIprefix\doi{10.1101/2020.07.14.187070}.
\bibitem[{Ma(2008)}]{ma2008dixon}
\bibinfo{author}{Ma, J.}, \bibinfo{year}{2008}.
\newblock \bibinfo{title}{Dixon techniques for water and fat imaging}.
\newblock \bibinfo{journal}{Journal of Magnetic Resonance Imaging}
  \bibinfo{volume}{28}, \bibinfo{pages}{543--558}.
\newblock \DOIprefix\doi{10.1002/jmri.21492}.
\bibitem[{McKay et~al.(2018)McKay, Wilman, Dennis, Kelly, Gyngell, Neubauer,
  Bell, Banerjee and Thomas}]{mckay2018measurement}
\bibinfo{author}{McKay, A.}, \bibinfo{author}{Wilman, H.R.},
  \bibinfo{author}{Dennis, A.}, \bibinfo{author}{Kelly, M.},
  \bibinfo{author}{Gyngell, M.L.}, \bibinfo{author}{Neubauer, S.},
  \bibinfo{author}{Bell, J.D.}, \bibinfo{author}{Banerjee, R.},
  \bibinfo{author}{Thomas, E.L.}, \bibinfo{year}{2018}.
\newblock \bibinfo{title}{Measurement of liver iron by magnetic resonance
  imaging in the {UK} {Biobank} population}.
\newblock \bibinfo{journal}{PLOS One} \bibinfo{volume}{13},
  \bibinfo{pages}{e0209340}.
\bibitem[{Mojtahed et~al.(2019)Mojtahed, Kelly, Herlihy, Kin, Wilman, McKay,
  Kelly, Milanesi, Neubauer, Thomas, Bell, Banerjee and
  Harisinghani}]{mojtahed2019reference}
\bibinfo{author}{Mojtahed, A.}, \bibinfo{author}{Kelly, C.J.},
  \bibinfo{author}{Herlihy, A.H.}, \bibinfo{author}{Kin, S.},
  \bibinfo{author}{Wilman, H.R.}, \bibinfo{author}{McKay, A.},
  \bibinfo{author}{Kelly, M.}, \bibinfo{author}{Milanesi, M.},
  \bibinfo{author}{Neubauer, S.}, \bibinfo{author}{Thomas, E.L.},
  \bibinfo{author}{Bell, J.D.}, \bibinfo{author}{Banerjee, R.},
  \bibinfo{author}{Harisinghani, M.}, \bibinfo{year}{2019}.
\newblock \bibinfo{title}{Reference range of liver corrected {T1} values in a
  population at low risk for fatty liver disease — a {UK} {Biobank}
  sub-study, with an appendix of interesting cases}.
\newblock \bibinfo{journal}{Abdominal Radiology} \bibinfo{volume}{44},
  \bibinfo{pages}{72--84}.
\bibitem[{{R Core Team}(2020)}]{r2020}
\bibinfo{author}{{R Core Team}}, \bibinfo{year}{2020}.
\newblock \bibinfo{title}{R: A Language and Environment for Statistical
  Computing}.
\newblock \bibinfo{organization}{R Foundation for Statistical Computing}.
  \bibinfo{address}{Vienna, Austria}.
\newblock \URLprefix \url{https://www.R-project.org}.
\bibitem[{Reeder et~al.(2005)Reeder, Pineda, Wen, Shimakawa, Yu, Brittain,
  Gold, Beaulieu and Pelc}]{reeder2005iterative}
\bibinfo{author}{Reeder, S.B.}, \bibinfo{author}{Pineda, A.R.},
  \bibinfo{author}{Wen, Z.}, \bibinfo{author}{Shimakawa, A.},
  \bibinfo{author}{Yu, H.}, \bibinfo{author}{Brittain, J.H.},
  \bibinfo{author}{Gold, G.E.}, \bibinfo{author}{Beaulieu, C.H.},
  \bibinfo{author}{Pelc, N.J.}, \bibinfo{year}{2005}.
\newblock \bibinfo{title}{Iterative decomposition of water and fat with echo
  asymmetry and least-squares estimation ({IDEAL}): Application with fast
  spin-echo imaging}.
\newblock \bibinfo{journal}{Magnetic Resonance in Medicine}
  \bibinfo{volume}{54}, \bibinfo{pages}{636--644}.
\bibitem[{Sudlow et~al.(2015)Sudlow, Gallacher, Allen, Beral, Burton, Danesh,
  Downey, Elliott, Green, Landray, Liu, Matthews, Ong, Pell, Silman, Young,
  Sprosen, Peakman and Collins}]{sudlow2015uk}
\bibinfo{author}{Sudlow, C.}, \bibinfo{author}{Gallacher, J.},
  \bibinfo{author}{Allen, N.}, \bibinfo{author}{Beral, V.},
  \bibinfo{author}{Burton, P.}, \bibinfo{author}{Danesh, J.},
  \bibinfo{author}{Downey, P.}, \bibinfo{author}{Elliott, P.},
  \bibinfo{author}{Green, J.}, \bibinfo{author}{Landray, M.},
  \bibinfo{author}{Liu, B.}, \bibinfo{author}{Matthews, P.},
  \bibinfo{author}{Ong, G.}, \bibinfo{author}{Pell, J.},
  \bibinfo{author}{Silman, A.}, \bibinfo{author}{Young, A.},
  \bibinfo{author}{Sprosen, T.}, \bibinfo{author}{Peakman, T.},
  \bibinfo{author}{Collins, R.}, \bibinfo{year}{2015}.
\newblock \bibinfo{title}{{UK} {Biobank}: An open access resource for
  identifying the causes of a wide range of complex diseases of middle and old
  age}.
\newblock \bibinfo{journal}{PLOS Medicine} \bibinfo{volume}{12},
  \bibinfo{pages}{e1001779}.
\bibitem[{Tarroni et~al.(2020)Tarroni, Bai, Oktay, Schuh, Suzuki, Glocker,
  Matthews and Rueckert}]{tarroni2020large}
\bibinfo{author}{Tarroni, G.}, \bibinfo{author}{Bai, W.},
  \bibinfo{author}{Oktay, O.}, \bibinfo{author}{Schuh, A.},
  \bibinfo{author}{Suzuki, H.}, \bibinfo{author}{Glocker, B.},
  \bibinfo{author}{Matthews, P.M.}, \bibinfo{author}{Rueckert, D.},
  \bibinfo{year}{2020}.
\newblock \bibinfo{title}{Large-scale quality control of cardiac imaging in
  population studies: Application to {UK} {Biobank}}.
\newblock \bibinfo{journal}{Scientific Reports} \bibinfo{volume}{10},
  \bibinfo{pages}{1--11}.
\bibitem[{Thomas et~al.(2013)Thomas, Fitzpatrick, Malik, Taylor-Robinson and
  Bell}]{thomas2013whole}
\bibinfo{author}{Thomas, E.L.}, \bibinfo{author}{Fitzpatrick, J.A.},
  \bibinfo{author}{Malik, S.J.}, \bibinfo{author}{Taylor-Robinson, S.D.},
  \bibinfo{author}{Bell, J.D.}, \bibinfo{year}{2013}.
\newblock \bibinfo{title}{Whole body fat: Content and distribution}.
\newblock \bibinfo{journal}{Progress in Nuclear Magnetic Resonance
  Spectroscopy} \bibinfo{volume}{73}, \bibinfo{pages}{56--80}.
\newblock \DOIprefix\doi{10.1016/j.pnmrs.2013.04.001}.
\bibitem[{Tustison et~al.(2010)Tustison, Avants, Cook, Zheng, Egan, Yushkevich
  and Gee}]{tustison2010n4itk}
\bibinfo{author}{Tustison, N.J.}, \bibinfo{author}{Avants, B.B.},
  \bibinfo{author}{Cook, P.A.}, \bibinfo{author}{Zheng, Y.},
  \bibinfo{author}{Egan, A.}, \bibinfo{author}{Yushkevich, P.A.},
  \bibinfo{author}{Gee, J.C.}, \bibinfo{year}{2010}.
\newblock \bibinfo{title}{{N4ITK}: Improved {N3} bias correction}.
\newblock \bibinfo{journal}{IEEE Transactions on Medical Imaging}
  \bibinfo{volume}{29}, \bibinfo{pages}{1310--1320}.
\bibitem[{{van Hout} et~al.(2020){van Hout}, Dekkers, Westenberg, Schalij,
  Scholte and Lamb}]{van2020impact}
\bibinfo{author}{{van Hout}, M.J.P.}, \bibinfo{author}{Dekkers, I.A.},
  \bibinfo{author}{Westenberg, J.J.M.}, \bibinfo{author}{Schalij, M.J.},
  \bibinfo{author}{Scholte, A.J.H.A.}, \bibinfo{author}{Lamb, H.J.},
  \bibinfo{year}{2020}.
\newblock \bibinfo{title}{The impact of visceral and general obesity on
  vascular and left ventricular function and geometry: a cross-sectional
  magnetic resonance imaging study of the {UK} {Biobank}}.
\newblock \bibinfo{journal}{European Heart Journal-Cardiovascular Imaging}
  \bibinfo{volume}{21}, \bibinfo{pages}{273--281}.
\bibitem[{Van~Rossum and Drake(2009)}]{python2009}
\bibinfo{author}{Van~Rossum, G.}, \bibinfo{author}{Drake, F.L.},
  \bibinfo{year}{2009}.
\newblock \bibinfo{title}{Python 3 Reference Manual}.
\newblock \bibinfo{publisher}{CreateSpace}, \bibinfo{address}{Scotts Valley,
  CA}.
\bibitem[{West et~al.(2016)West, {Dahlqvist Leinhard}, Romu, Collins, Garratt,
  Bell, Borga and Thomas}]{west2016feasibility}
\bibinfo{author}{West, J.}, \bibinfo{author}{{Dahlqvist Leinhard}, O.},
  \bibinfo{author}{Romu, T.}, \bibinfo{author}{Collins, R.},
  \bibinfo{author}{Garratt, S.}, \bibinfo{author}{Bell, J.D.},
  \bibinfo{author}{Borga, M.}, \bibinfo{author}{Thomas, E.L.},
  \bibinfo{year}{2016}.
\newblock \bibinfo{title}{Feasibility of {MR}-based body composition analysis
  in large scale population studies}.
\newblock \bibinfo{journal}{PLOS One} \bibinfo{volume}{11},
  \bibinfo{pages}{e0163332}.
\bibitem[{Wilman et~al.(2017)Wilman, Kelly, Garratt, Matthews, Milanesi,
  Herlihy, Gyngell, Neubauer, Bell, Banerjee and
  Thomas}]{wilman2017characterisation}
\bibinfo{author}{Wilman, H.R.}, \bibinfo{author}{Kelly, M.},
  \bibinfo{author}{Garratt, S.}, \bibinfo{author}{Matthews, P.M.},
  \bibinfo{author}{Milanesi, M.}, \bibinfo{author}{Herlihy, A.},
  \bibinfo{author}{Gyngell, M.}, \bibinfo{author}{Neubauer, S.},
  \bibinfo{author}{Bell, J.D.}, \bibinfo{author}{Banerjee, R.},
  \bibinfo{author}{Thomas, E.L.}, \bibinfo{year}{2017}.
\newblock \bibinfo{title}{Characterisation of liver fat in the {UK} {Biobank}
  cohort}.
\newblock \bibinfo{journal}{PLOS One} \bibinfo{volume}{12},
  \bibinfo{pages}{e0172921}.
\bibitem[{Wilman et~al.(2019)Wilman, Parisinos, Atabaki-Pasdar, Kelly, Thomas,
  Neubauer, {IMI DIRECT Consortium}, Mahajan, Hingorani, Patel, Hemingway,
  Franks, Bell, Banerjee and Yaghootkar}]{wilman2019genetic}
\bibinfo{author}{Wilman, H.R.}, \bibinfo{author}{Parisinos, C.A.},
  \bibinfo{author}{Atabaki-Pasdar, N.}, \bibinfo{author}{Kelly, M.},
  \bibinfo{author}{Thomas, E.L.}, \bibinfo{author}{Neubauer, S.},
  \bibinfo{author}{{IMI DIRECT Consortium}}, \bibinfo{author}{Mahajan, A.},
  \bibinfo{author}{Hingorani, A.D.}, \bibinfo{author}{Patel, R.S.},
  \bibinfo{author}{Hemingway, H.}, \bibinfo{author}{Franks, P.W.},
  \bibinfo{author}{Bell, J.D.}, \bibinfo{author}{Banerjee, R.},
  \bibinfo{author}{Yaghootkar, H.}, \bibinfo{year}{2019}.
\newblock \bibinfo{title}{Genetic studies of abdominal {MRI} data identify
  genes regulating hepcidin as major determinants of liver iron concentration}.
\newblock \bibinfo{journal}{Journal of Hepatology} \bibinfo{volume}{71},
  \bibinfo{pages}{594--602}.
\bibitem[{Wojciechowska et~al.(2018)Wojciechowska, Irving, Dennis, Wilman,
  Banerjee, Brady and Kelly}]{wojciechowska2018automated}
\bibinfo{author}{Wojciechowska, M.}, \bibinfo{author}{Irving, B.},
  \bibinfo{author}{Dennis, A.}, \bibinfo{author}{Wilman, H.R.},
  \bibinfo{author}{Banerjee, R.}, \bibinfo{author}{Brady, M.},
  \bibinfo{author}{Kelly, M.}, \bibinfo{year}{2018}.
\newblock \bibinfo{title}{Automated detection of cystic lesions in quantitative
  {T1} liver images}, in: \bibinfo{booktitle}{Annual Conference on Medical
  Image Understanding and Analysis}, pp. \bibinfo{pages}{51--56}.
\bibitem[{Wood et~al.(2005)Wood, Enriquez, Ghugre, Tyzka, Carson, Nelson and
  Coates}]{wood2005mri}
\bibinfo{author}{Wood, J.C.}, \bibinfo{author}{Enriquez, C.},
  \bibinfo{author}{Ghugre, N.}, \bibinfo{author}{Tyzka, J.M.},
  \bibinfo{author}{Carson, S.}, \bibinfo{author}{Nelson, M.D.},
  \bibinfo{author}{Coates, T.D.}, \bibinfo{year}{2005}.
\newblock \bibinfo{title}{{MRI} {R2} and {R2*} mapping accurately estimates
  hepatic iron concentration in transfusion-dependent thalassemia and sickle
  cell disease patients}.
\newblock \bibinfo{journal}{Blood} \bibinfo{volume}{106},
  \bibinfo{pages}{1460--1465}.
\bibitem[{Yu et~al.(2008)Yu, Shimakawa, McKenzie, Brodsky, Brittain and
  Reeder}]{yu2008multiecho}
\bibinfo{author}{Yu, H.}, \bibinfo{author}{Shimakawa, A.},
  \bibinfo{author}{McKenzie, C.A.}, \bibinfo{author}{Brodsky, E.},
  \bibinfo{author}{Brittain, J.H.}, \bibinfo{author}{Reeder, S.B.},
  \bibinfo{year}{2008}.
\newblock \bibinfo{title}{Multiecho water-fat separation and simultaneous {R2*}
  estimation with multifrequency fat spectrum modeling}.
\newblock \bibinfo{journal}{Magnetic Resonance in Medicine}
  \bibinfo{volume}{60}, \bibinfo{pages}{1122--1134}.

\end{thebibliography}


\end{document}